\documentclass[twocolumn,prx,aps,showpacs,floatfix,superscriptaddress,preprintnumbers,amsmath,amssymb]{revtex4}

\usepackage{graphicx}
\usepackage{bbm}
\usepackage{amsbsy}

\begin{document}

\newcommand{\mfs}{Mn$_{1-x}$Fe$_{x}$Si}
\newcommand{\mcs}{Mn$_{1-x}$Co$_{x}$Si}
\newcommand{\fcs}{Fe$_{1-x}$Co$_{x}$Si}
\newcommand{\cso}{Cu$_{2}$OSeO$_{3}$}

\newcommand{\rxx}{$\rho_{\rm xx}$}
\newcommand{\rxy}{$\rho_{\rm xy}$}
\newcommand{\rxytop}{$\rho_{\rm xy}^{\rm top}$}
\newcommand{\Drxyt}{$\Delta\rho_{\rm xy}^{\rm top}$}
\newcommand{\Sxy}{$\sigma_{\rm xy}$}
\newcommand{\Sxya}{$\sigma_{\rm xy}^{A}$}

\newcommand{\bco}{$B_{\rm c1}$}
\newcommand{\bct}{$B_{\rm c2}$}
\newcommand{\bao}{$B_{\rm A1}$}
\newcommand{\bat}{$B_{\rm A2}$}
\newcommand{\beff}{$B^{\rm eff}$}

\newcommand{\tc}{$T_{\rm c}$}

\newcommand{\mb}{$\mu_0\,M/B$}
\newcommand{\dmdb}{$\mu_0\,\mathrm{d}M/\mathrm{d}B$}
\newcommand{\ddmddb}{$\mathrm{\mu_0\Delta}M/\mathrm{\Delta}B$}
\newcommand{\cm}{$\chi_{\rm M}$}
\newcommand{\cac}{$\chi_{\rm ac}$}
\newcommand{\rechi}{${\rm Re}\,\chi_{\rm ac}$}
\newcommand{\imchi}{${\rm Im}\,\chi_{\rm ac}$}

\newcommand{\ozz}{$\langle100\rangle$}
\newcommand{\ooz}{$\langle110\rangle$}
\newcommand{\ooo}{$\langle111\rangle$}
\newcommand{\too}{$\langle211\rangle$}

\renewcommand{\vec}[1]{{\bf #1}}

\title{Giant generic topological Hall resistivity of MnSi under pressure}

\author{R. Ritz}
\affiliation{Technische Universit\"at M\"unchen, Physik-Department E21, D-85748 Garching, Germany}

\author{M. Halder}
\affiliation{Technische Universit\"at M\"unchen, Physik-Department E21, D-85748 Garching, Germany}

\author{C. Franz}
\affiliation{Technische Universit\"at M\"unchen, Physik-Department E21, D-85748 Garching, Germany}

\author{A. Bauer}
\affiliation{Technische Universit\"at M\"unchen, Physik-Department E21, D-85748 Garching, Germany}

\author{M. Wagner}
\affiliation{Technische Universit\"at M\"unchen, Physik-Department E21, D-85748 Garching, Germany}

\author{R. Bamler}
\affiliation{Institute of Theoretical Physics, Universit\"at zu K\"oln, D-50937 K\"oln, Germany}

\author{A. Rosch}
\affiliation{Institute of Theoretical Physics, Universit\"at zu K\"oln, D-50937 K\"oln, Germany}

\author{C. Pfleiderer}
\affiliation{Technische Universit\"at M\"unchen, Physik-Department E21, D-85748 Garching, Germany}

\date{\today}

\begin{abstract}
We report detailed low temperature magnetotransport and magnetization measurements in MnSi under pressures up to $\sim12\,{\rm kbar}$. Tracking the role of sample quality, pressure transmitter, and field and temperature history allows us to link the emergence of a giant topological Hall resistivity $\sim50\,{\rm n\Omega cm}$ to the skyrmion lattice phase at ambient pressure. We show that the remarkably large size of the topological Hall resistivity in the zero-temperature limit  must be generic. We discuss various mechanisms which can lead to the much smaller signal at elevated temperatures observed at ambient pressure. 
\end{abstract}

\pacs{75.25-j, 75.50.-y, 75.10-b}

\vskip2pc

\maketitle

\section{Motivation}
Changes of the Berry phase of the conduction electrons in metals reflect sensitively the topology of adiabatic changes of their spin orientation. These Berry phases can be described by emergent (fictitious) Aharonov-Bohm magnetic fields $B^{\rm eff}$ which control the quasiclassical motion of electrons in phase space.  As the Aharonov-Bohm fields cause a deflection of the trajectory in the plane perpendicular to the field, they lead to a Hall signal. Two limits of this Berry-phase deflection may be distinguished \cite{Nagaosa:RMP10,Xiao:RMP10}. 

On the one hand,  spin-orbit coupling and local electric fields varying on an atomic length scale may lead to a band structure where the spin orientation depends on its momentum. The resulting Berry phases can be described by an emergent magnetic field which acts, however, not in position but in momentum space. In this case, an anomalous contribution to the Hall conductivity {\Sxya} arises in terms of dissipationless Hall currents, which reflect differences of the Berry phase collected by majority and minority charge carriers. In the simplest scenario, {\Sxya} scales with the uniform spin polarization. This is referred to as the \textit{intrinsic anomalous} Hall effect since  {\Sxya} turns out to be independent of impurity scattering. In turn, this implies that the corresponding Hall resistivity $\rho_{\rm xy}$ is proportional to the square of the longitudinal resistivity, $\rho_{\rm xy}\approx - \sigma^A_{\rm xy} \rho_{\rm xx}^2$, where we assumed $\sigma_{\rm xy} \ll \sigma_{\rm xx}$ as for most good metals.

On the other hand, smoothly varying magnetic textures, which change their spin orientation on length scales much longer than the Fermi wavelength, give rise to Berry phases picked up in real space. The corresponding emergent magnetic field acts similar to a real magnetic field. This effect may be described in terms of quasiparticles supporting emergent charges with the important difference that majority and minority electrons carry opposite emergent charges (a technical description follows). As for smooth magnetic textures, the real-space Berry phases are directly associated to the real-space winding of the magnetization, the corresponding contribution to the Hall effect is then referred to as the \textit{topological} Hall resistivity,  \rxytop. As for the conventional Hall effect,  {\rxytop} is approximately independent of the total scattering rate. In multiband systems, however, the relative strength of scattering rates determines the relative size of contributions from the various bands and therefore also the size of \rxytop.

The concepts of real- and momentum-space Berry phases may also be generalized to Berry phases in phase space \cite{Xiao:RMP10}. These arise when the local direction of the electron spin is governed both by  spin-orbit coupling in the bands and by smooth magnetic textures. Their importance in real materials is essentially unexplored.


An increasing number of experimental studies support the existence of the intrinsic anomalous Hall effect \cite{Nagaosa:RMP10,Lee:Science2004}. A vital piece of evidence is thereby related to the temperature dependence of the Hall conductivity, which scales with the magnetization. In turn, the intrinsic anomalous Hall resistivity vanishes for $T\to0$ in high-purity metals with low residual resistivities as $\rho_{\rm xy}^A \propto \rho^2$. This is contrasted by the temperature and field dependence of the topological Hall resistivity, which is approximately independent of the elastic scattering rate. As minority and majority electrons carry opposite emergent charges, the topological Hall resistivity is thereby sensitive to the strength of the local magnetization. Since the difference in density of minority and majority electrons also decreases with increasing temperature, {\rxytop} may therefore be expected to decrease with increasing temperature in materials with a well-defined nonzero topological winding number per magnetic unit cell. Qualitatively, the temperature dependence caused by the local magnetization may be enhanced by spin-flip scattering, which prohibits that the electrons follow the magnetic texture adiabatically. Since spin-flip scattering typically increases with increasing temperature the topological Hall signal may therefore decrease even faster than expected from the temperature dependence of the  difference of majority and minority charge carriers alone. 

Numerous experimental studies have addressed the existence of topological Hall contributions. However, their  identification has been ambiguous, especially in the absence of topological quantization, i.e., nonzero topological winding of the spin structure per magnetic unit cell. For instance, in a seminal paper, a topological Hall signal has been reported for three-dimensional pyrochlore lattices \cite{Taguchi:Science01,Machida:PRL07}. Yet, in these systems the noncoplanar spin structure is due to geometric frustration on short length scales, which thus cannot be described as a smoothly varying structure in position space. Moreover, the topological Hall effect is not related to a nonzero topological winding number per magnetic unit cell. Recently, an analogous study reported a large topological Hall signal for geometrically frustrated noncoliniear spin order in UCu$_5$ \cite{Ueland:NC2012}. Likewise, a topological Hall signal has also been considered, e.g., in La$_{1-x}$Co$_x$MnO$_3$ \cite{Ye:PRL99}, CrO$_2$ \cite{Yanagihara:PRL02}, and Gd \cite{Baily:PRB05}. Yet, for these systems there is essentially no independent microscopic information on the relevant spin structures.  

A new generation of experimental studies of the topological Hall effect has become possible with the discovery of lattices of magnetic whirl lines, so-called skyrmions, in chiral magnets such as MnSi, described in further detail in Sec.\,\ref{mnsi}. The skyrmion lattice represents the first example of long-range magnetic order with a well-defined nonzero topological winding number per magnetic unit cell. Measurements of the Hall effect have indeed revealed a contribution to the Hall signal, which appeared to be switched on and off when entering or leaving the skyrmion phase, respectively \cite{Neubauer:PRL2009}. This was attributed to the topological Hall effect. In a pioneering high-pressure study of Lee and co-workers \cite{Lee:PRL09}, the existence of a tenfold larger, hitherto unexplained, topological Hall signal was reported above 6\,kbar, where no data was shown between ambient pressure and 6\,kbar. The main goal of the work reported in our paper is to clarify how the large signal reported in Ref.\,\cite{Lee:PRL09} is related to the magnetic phase diagram of MnSi and the skyrmion lattice phase at ambient pressure and which factors determine the size of the Hall signal.

Clarification of the origin of the large topological Hall signal observed in MnSi under pressure provides an important point of reference for a wide range of problems. First, recent experiments have identified the effects of spin-transfer torques in the skyrmion lattice of MnSi at tiny electric current densities \cite{Jonietz:Science2010,Schulz:NaturePhysics2012}. Here, the size of the topological Hall signal reflects the strength of the coupling between the electric currents and the spin structure. Thus, understanding the size of the topological Hall effect promises major advances in the understanding of the origin of spin-transfer torques. Second, the largest topological Hall signals have so far been reported for MnGe (Ref.\,\cite{Kanzawa:PRL2011}) and SrFeO$_3$ (Ref.\,\cite{Ishiwata:PRB2011}), reaching up to $\sim200\,{\rm n\Omega cm}$. It has been speculated that this provides evidence for Aharanov-Bohm fields up to many hundred tesla. If correct, this might pave the way to a completely new generation of phenomena in which even larger emergent fields approach the quantum limit. It is therefore of great interest to gain an understanding as to what determines the quantitative size of the topological Hall signal and whether the generic size of the topological Hall effect may be even much larger. Third, understanding the topological Hall signals in skyrmion lattices and related structures will also shed new light on the large number of more conventional materials in which topological Hall effects have been claimed.

\begin{figure}
\includegraphics[width=0.3\textwidth]{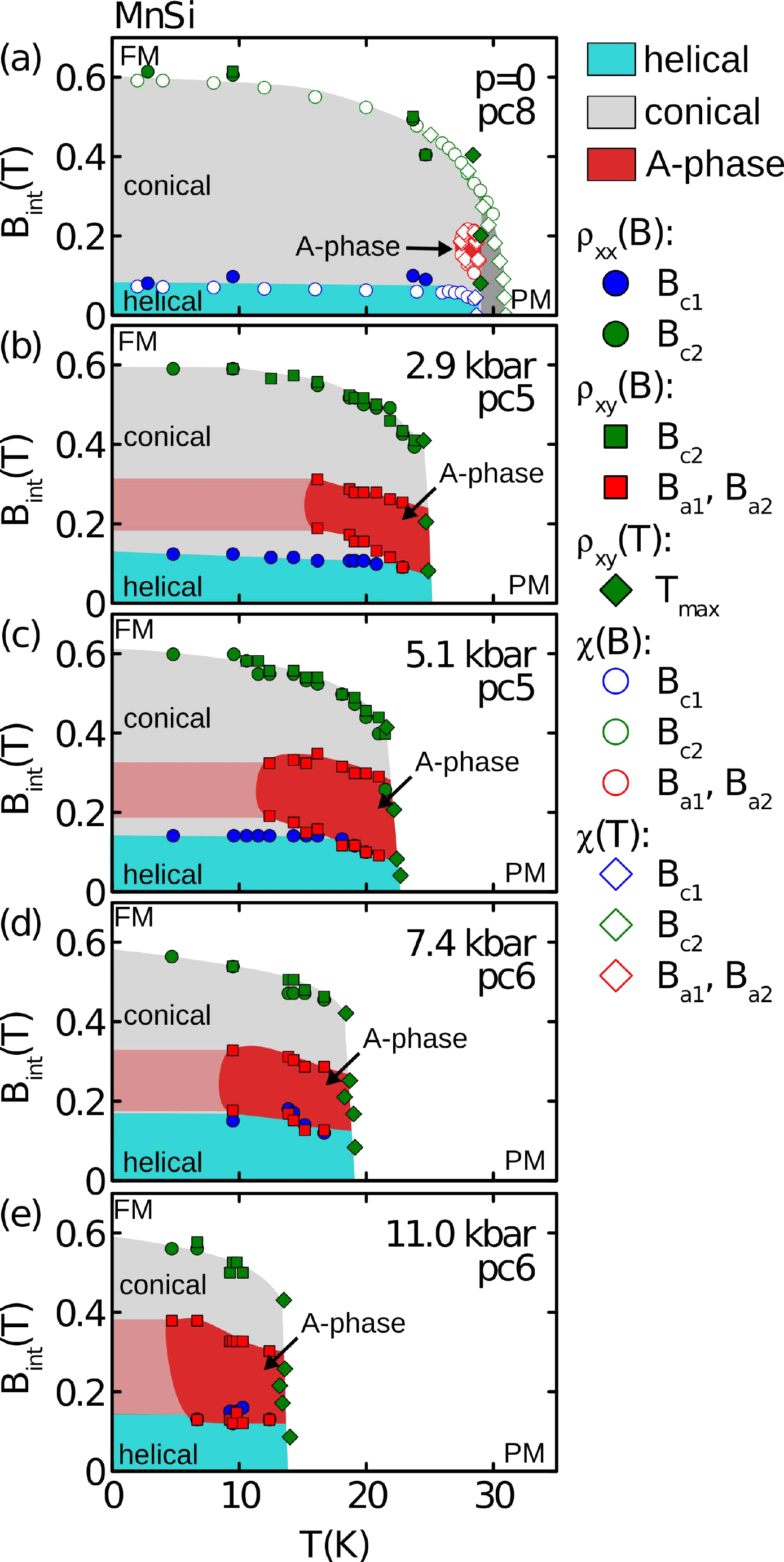}
\caption{(Color online) Magnetic phase diagrams of MnSi at various pressures inferred from the magnetotransport properties for $B\parallel \langle110\rangle$. Plots are based on data recorded with a methanol:ethanol (ME) mixture as pressure transmitter. Dark red shading, denoted as the A phase, represents the regime of an additional topological Hall signal in field scans arising from the skyrmion lattice. Bright red shading indicates the metastable topological Hall signal under field cooling.
}
\label{figure7}
\end{figure}

\begin{figure}
\includegraphics[width=0.3\textwidth]{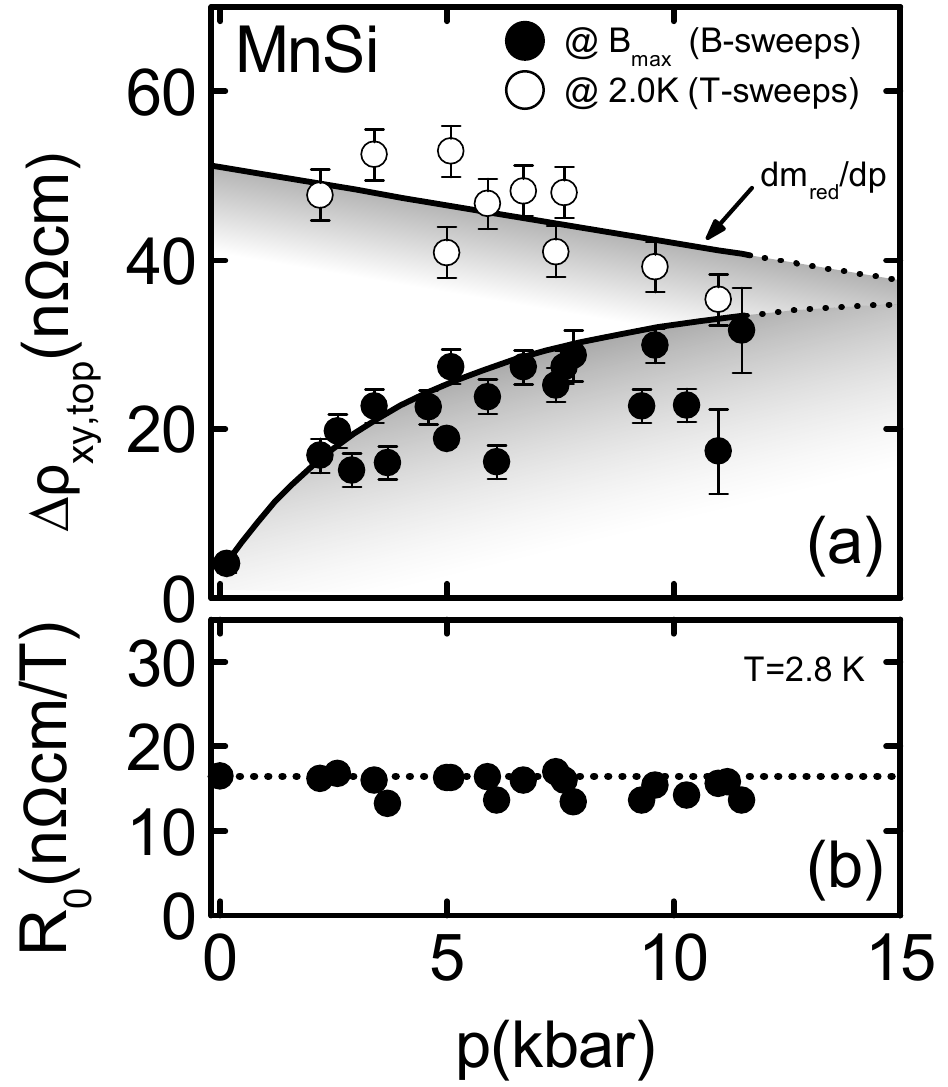}
\caption{Estimated magnitude of the topological contribution to the Hall effect and the normal Hall constant $R_0$ as a function of pressure. (a) Full symbols represent the maximum topological Hall contribution observed in field sweeps slightly below {\tc}. Open symbols represent a metastable topological Hall contribution under field cooling for temperatures of 2\,K [for clarity, only data from Fig.\,\ref{figure4}\,(f) are taken into account here]. (b) Normal Hall constant determined at 2.8\,K as the slope of {\rxy} at large fields around 10\,T.
}
\label{figure10}
\end{figure}

In this paper, we address the question of the generic size of the topological Hall effect in terms of a comprehensive high-pressure study of the itinerant electron magnet MnSi. For ease of reading, we summarize in Figs.\,\ref{figure7} and \ref{figure10} our main results consisting in the evolution of the magnetic phase diagram and the topological Hall signal as a function of pressure, respectively. Regimes in the magnetic phase diagrams, where we observe a topological Hall signal, are thereby shown in red shading, where dark red shading refers to reversible behavior in field sweeps as well as zero-field cooling and field cooling. An important aspect of our study is the additional discovery that the topological Hall signal for pressures larger than ambient pressure survives under field cooling down to the lowest temperatures. This is illustrated in Fig.\,\ref{figure7} by light red shading. Based on the dependence on field and temperature history, we obtain an estimate of the size of the topological Hall signal and its pressure dependence in the zero-temperature limit as shown in Fig.\,\ref{figure10}. Here, full symbols represent the maximum topological Hall contribution observed in field sweeps slightly below {\tc}. Open symbols represent a metastable topological Hall contribution under field cooling for temperatures of 2\,K, the lowest temperature measured, i.e., the topological Hall signal without the degrading effects of finite temperatures. Taken together, our study establishes that the emergence of a giant generic topological Hall signal under pressure is connected with the skyrmion lattice phase at ambient pressure. We thereby identify pressure inhomogeneities and mediocre sample quality [low residual resistivity ratios (RRRs)] as important factors that influenced previous high-pressure studies. We further identify that  temperature is an important factor affecting the size of the topological Hall signal and discuss various mechanisms which explain its strong variation with pressure.

The presentation of our study is organized as follows. We continue our introduction with the properties of MnSi in Sec.\,\ref{mnsi}, describing in detail the understanding of the skyrmion lattice phase and the topological Hall signal as reported so far. This is followed in Sec.\,\ref{methods} by an account of the experimental methods, where we specifically address the role of the pressure transmitter, cooling conditions, sample quality, and temperature and field history. The presentation of our results in Sec.\,\ref{results} begins with the magnetic field dependence of both the electrical transport properties and the magnetization, followed by their temperature dependence. This allows to appreciate better the metastable properties we observe under field cooling. The brief theoretical discussion in Sec.\,\ref{theory} focuses on the interplay of topological and anomalous Hall effect. The paper concludes with a discussion of the results observed in Sec.\,\ref{discussion}, where we consider various factors which determine the size of the anomalous Hall effect. A short set of conclusions is given in Sec.\,\ref{conclusions}. 

\section{Introduction to MnSi}
\label{mnsi}

The itinerant-electron magnet MnSi is ideally suited to pursue the question of the generic size of the topological Hall resistivity in a real material. In the noncentrosymmetric cubic $B20$ crystal structure, space group $P2_13$, three hierarchical energy scales account for the magnetic properties \cite{Landau}. A competition between ferromagnetic exchange and Dzyaloshinsky-Moriya interactions on the strongest and second strongest scales, respectively, generates a long-wavelength helimagnetic modulation, $\lambda_{\rm h} (T\to0)\sim\,180\,{\rm \AA}$, below $T_{\rm c}\approx29.5\,{\rm K}$ \cite{wavelength}. The helical modulation propagates along the cubic space diagonal {\ooo} due to magnetic anisotropies by higher-order spin-orbit coupling providing the weakest scale.

Of particular interest is the magnetic phase diagram of MnSi. Below $T_c$ and as a function of increasing magnetic field, the helimagnetic order undergoes a reorientation transition at a field $B_{\rm c1}\approx0.1\,{\rm T}$ into a spin-flop phase also known as conical phase. Depending on field direction, this reorientation is either a crossover or a symmetry-breaking second-order phase transition. When further increasing the field, a second transition takes place from the conical phase to a spin polarized (ferromagnetic) state at $B_{\rm c2}\approx0.6\,{\rm T}$ \cite{Kusaka:SSC1976,Kadowaki:JPSJ1982,Bauer:PRB12}.  Finally, in the vicinity of $T_c$, a small additional phase pocket exists within the conical phase, historically referred to as the A phase \cite{Kadowaki:JPSJ1982}. Although the A phase had been known for a long time, the underlying spin structure was only recently identified, providing the first example of long-range magnetic order in which each magnetic unit cell supports a nonzero topological winding number \cite{Muehlbauer:Science2009,Adams:PRL2011,Neubauer:PRL2009,Tonumura:NanoLetters2012}. The skyrmions may be visualized as a kind of vortex lines that stabilize parallel to the applied magnetic field. 

The topological winding number supported by the skyrmion lattice may be determined by integrating the winding density over the two-dimensional magnetic unit cell (UC):
$\Phi=\int_{\rm UC}\mathbf \Phi\, {\bf d}^2 \bf r$, 
where  $\Phi^\mu=\frac{1}{8\pi} \epsilon_{\rm \mu\nu\lambda}
\hat{n}\cdot(\partial_{\rm \nu}\hat{n}\times\partial_{\rm \lambda}\hat{n})$. 
Here, $\epsilon_{\rm \mu\nu\lambda}$ is the antisymmetric unit tensor and 
$\hat{n}=\vec{M}/\vert M \vert$ the magnetic unit vector \cite{comment-machida07}. For the case of the skyrmion lattice in MnSi, a winding number $\Phi=-1$ per magnetic unit cell is obtained. Thus, in contrast to some of the noncollinear magnetic structures in geometrically frustrated magnets, where the winding senses are staggered causing cancellations of topological contributions to the Hall resistivity, no such cancellations occur in the skyrmion lattice of chiral magnets since the topological winding is quantized and nonzero per magnetic unit cell. 

Measurements of the Hall effect in MnSi have revealed three contributions: first, the normal Hall effect $\rho_{\rm n}=R_0\,B$, which is proportional to the applied magnetic field $B$; second, an intrinsic anomalous Hall conductivity, $\sigma_{\rm xy}$, below the helimagnetic transition temperature, $T_{\rm c}=29.5\,{\rm K}$, which scales with the magnetization $\sigma_{\rm xy}=S_{\rm H}\,M$ \cite{Lee:PRB07,Neubauer:PhysicaB09,clarification}; third, a topological Hall signal, $\rho_{\rm xy}^{\rm top}$, in the regime of the skyrmion lattice phase, which reflects the nonzero topological winding number of the spin structure. Assuming the absence of intraband (spin-flip) scattering and that the interband (non-spin-flip) scattering may be captured by the normal Hall constant $R_0$, an estimate of {\rxytop} has been given by $\rho_{\rm xy}^{\rm top}=P\,R_0\,B^{\rm eff}$. Here, $P$ is the charge carrier spin polarization, and $B^{\rm eff}$ the emergent Aharonov-Bohm field associated with the Berry phase arising from the topological winding of the texture \cite{Neubauer:PRL2009} (see Sec.\,\ref{theory} for an account of the considerations entering this formula and for a precise definition of $P$). The emergent magnetic field per magnetic unit cell is thereby topologically quantized \cite{Ye:PRL99,Bruno:PRL04,Tatara:JPSJ07,Binz:PhysicaB08}: its average strength is given by one flux quantum per magnetic unit cell times the winding number (see Sec.\,\ref{theory}). 

For a quantitative estimate, we note that the skyrmion lattice is hexagonal, similar to the vortex lattice in type-II superconductors. Hence, the reciprocal and real-space lattice vectors of the skyrmion lattice have length $2 \pi/\lambda_{\rm S} $ and $\lambda_{\rm S} /\sin(2 \pi/3)$, respectively, where $\lambda_{\rm S}$ corresponds approximately to the wavelength of the helical state near $T_c$, $\lambda_{\rm h} \approx \lambda_{\rm S} \approx 165\,{\rm \AA}$  \cite{wavelength}. The size of the unit cell is in turn given by $\lambda_{\rm S}^2/\sin(2 \pi/3)$ and one obtains for MnSi
\begin{eqnarray}\label{beff}
B^{\rm eff}
=- \frac{h}{e} \left( \frac{\sqrt 3}{2 \lambda_{\rm S}
^2}\right) \approx
-13.15 \,{\rm T}.
\end{eqnarray}
The sign thereby reflects the winding number of $-1$, implying that the emergent field $B^{\rm eff}$ is oriented \textit{antiparallel} to the physical magnetic field. 

We note that the same expression for $B^{\rm eff}$ of minus one flux quantum per unit cell was used in Ref.~\cite{Neubauer:PRL2009}. However, due to an unfortunate calculational mistake, a five times smaller value for $B^{\rm eff}\approx-2.5\,{\rm T}$ was stated in this paper \cite{erratum}. The  experimentally observed~\cite{Neubauer:PRL2009} contribution $\rho_{\rm xy}^{\rm top}\approx -(4\pm1)\,{\rm n\Omega cm}$ is hence approximately a factor of 5 smaller than the theoretically estimated value inferred from the simple expression $\rho_{\rm xy}^{\rm top}=P\,R_0\,B^{\rm eff}$, when one uses $P\approx0.1$ and $R_0=1.7\cdot10^{-10}\,{\rm \Omega m T^{-1}}$ as in Ref.~\cite{Neubauer:PRL2009}. Both $P$ and $R_0$ thereby yield considerable uncertainties. As discussed in Sec.\,\ref{theory},  $P$ depends on a complicated Fermi-surface average, while the value of $R_0$ is difficult to extract in the relevant parameter regime due to a huge anomalous Hall contribution. For example, the normal Hall constant $R_0$, given above, which was used for the estimate in Ref.\,\cite{Neubauer:PRL2009}, was inferred from the Hall signal at room temperature, while a simultaneous fit of the normal and anomalous Hall signal at low temperatures suggests $R_0\approx-0.8\cdot10^{-10}\,{\rm \Omega m T^{-1}}$ \cite{Lee:PRB07,Neubauer:PhysicaB09}. Nevertheless, contrary to the conclusions of Ref.\,\cite{Neubauer:PRL2009}, the value of $\rho_{\rm xy}^{\rm top}$ observed experimentally is, in fact, much smaller than the theoretical prediction. This may be explained, in principle, by several mechanisms discussed in detail below, which were not considered in Ref.\,\cite{Neubauer:PRL2009}.

The detailed pressure dependence of the helimagnetic properties of MnSi offer a fresh perspective as concerns the origin and the size of the topological Hall signal. In a pioneering study, Lee \textit{et al.} reported a topological Hall signal for hydrostatic pressures in the range 6 to 12\,kbar (Ref.~\cite{Lee:PRL09}) that seemed puzzling in two ways. First, the signal was very large, $\rho_{\rm xy}^{\rm top}\sim -40\,{\rm n\Omega cm}$. Such a tenfold larger signal either requires a drastic reduction of the skyrmion lattice spacing by over a factor of 3, or a tenfold increase of the conduction spin polarization $P$ or a tendfold increase of the normal Hall constant $R_0$ (or a highly unusual combination of these aspects). Second, the field range of this very large topological Hall signal extended all the way from {\bco} to {\bct} and did not correspond to the range over which a skyrmion lattice phase may be expected based on the phase diagram at ambient pressure.  

The results reported by Lee \textit{et al.} \cite{Lee:PRL09} and their relationship to the magnetic phase diagram at ambient pressure seemed also surprising in view of the very detailed high-pressure studies of MnSi reported in the literature. These comprise measurements of the resistivity \cite{Pfleiderer:PRB97,Thessieu:SSC95,Doiron:Nature03,Petrova:PRB2012}, ac susceptibility \cite{Thessieu:JPCM97}, magnetization \cite{Bloch:PLA74,Thessieu:JPSJ98,Koyama:PRB00}, thermal expansion \cite{Pfleiderer:Science07}, thermopower \cite{Cheng:PRB10}, neutron scattering \cite{Pfleiderer:Nature2004,Pfleiderer:PRL07}, NMR \cite{Yu:PRL04}, and $\mu$-SR \cite{Uemura:NaturePhysics07}. They may be summarized as follows \cite{comment-stishov}. With increasing pressure, the helimagnetic transition temperature measured in the resistivity, ac susceptibility, and magnetization decreases and vanishes around $p_c\approx14.6\,{\rm kbar}$. The temperature-versus-pressure phase diagram displays considerable complexities for $p>p^*\sim{\rm 12\,kbar}$. For instance, the helimagnetic transition turns distinctly first order, where the appearance of itinerant metamagnetism under applied magnetic fields provides the most striking evidence. Neutron scattering, $\mu$-SR, and NMR  suggest phase separation of the magnetic order for $p>p^*\sim{\rm 12\,kbar}$, where a decreasing volume fraction of helimagnetic order tracks  $T_c(p)$ as inferred from the resistivity and ac susceptibility. 

A major puzzle surrounds the temperature dependence of the electrical resistivity which displays a $T^{3/2}$ non-Fermi-liquid form for $T\lesssim 12\,{\rm K}$ in the regime where the helimagnetic order has been suppressed, namely $p>p_c$. The non-Fermi-liquid resistivity thereby survives up to pressures of at least $2\,p_c$, contrasting the expectations of a conventional quantum critical point. Neutron scattering moreover shows the presence of a peculiar magnetic scattering intensity on the surface of a small sphere in reciprocal space at a wavelength of the helical modulation. By analogy with liquid crystals, this scattering pattern has been referred to as partial order. As the partial order extends deep into the NFL regime without signs of phase transitions, it appears to be the signature of a spin liquid, possibly with nontrivial topological character. 

Regarding the possible origin of the very large topological Hall signal, reported in the pressure range from 6 to 12\,kbar, studies reported in the literature provide the following key information. First, the wavelength of the helical modulation is essentially unchanged as a function of pressure. This implies that a skyrmion lattice phase associated with the helimagnetic state, like that at ambient pressure, has an unchanged lattice constant, i.e., $B^{\rm eff}$ remains essentially unchanged under pressure. Second, the ordered magnetic moment in the helimagnetic state decreases gently by $\sim10\,\%$ up to 12\,kbar (it certainly does not increase). Since the Curie-Weiss moment in the paramagnetic state is also unchanged, the polarization $P$ of the electron bands does not appear to change drastically. Third, neither the electrical resistivity, which is well behaved, nor the normal Hall effect, as inferred from the data shown by Lee \textit{et al.} suggest a change of $R_0$ as a function of pressure (Lee \textit{et al.} do not comment on the pressure dependence of $R_0$). Finally, the critical fields {\bco} and {\bct} do not change under pressure, consistent with the unchanged helical modulation. Hence, the phase boundaries of the skyrmion lattice phase should be unchanged. In turn, the size and the field range of the topological Hall signal reported by Lee \textit{et al.} either represents a completely novel phenomenon or an unexpected conspiracy of mechanisms, both of which are of great interest.

In order to identify the generic size of the topological Hall signal in a well-known material, we have revisited the pressure dependence of MnSi reported in Ref.\,\cite{Lee:PRL09} and its seeming inconsistencies with the topological Hall signal at ambient pressure \cite{Neubauer:PRL2009}. In the study reported here we present data up to $p^*\approx12\,{\rm kbar}$, avoiding the complexities associated with the first-order transition, itinerant metamagnetism, phase coexistence, partial magnetic order, and extended non-Fermi-liquid resistivity. As all of these aspects are beyond the scope of the work presented here, Hall effect measurements above $p^*$ will be reported elsewhere \cite{Ritz:Nature2013}.

\begin{figure}
\includegraphics[width=0.35\textwidth]{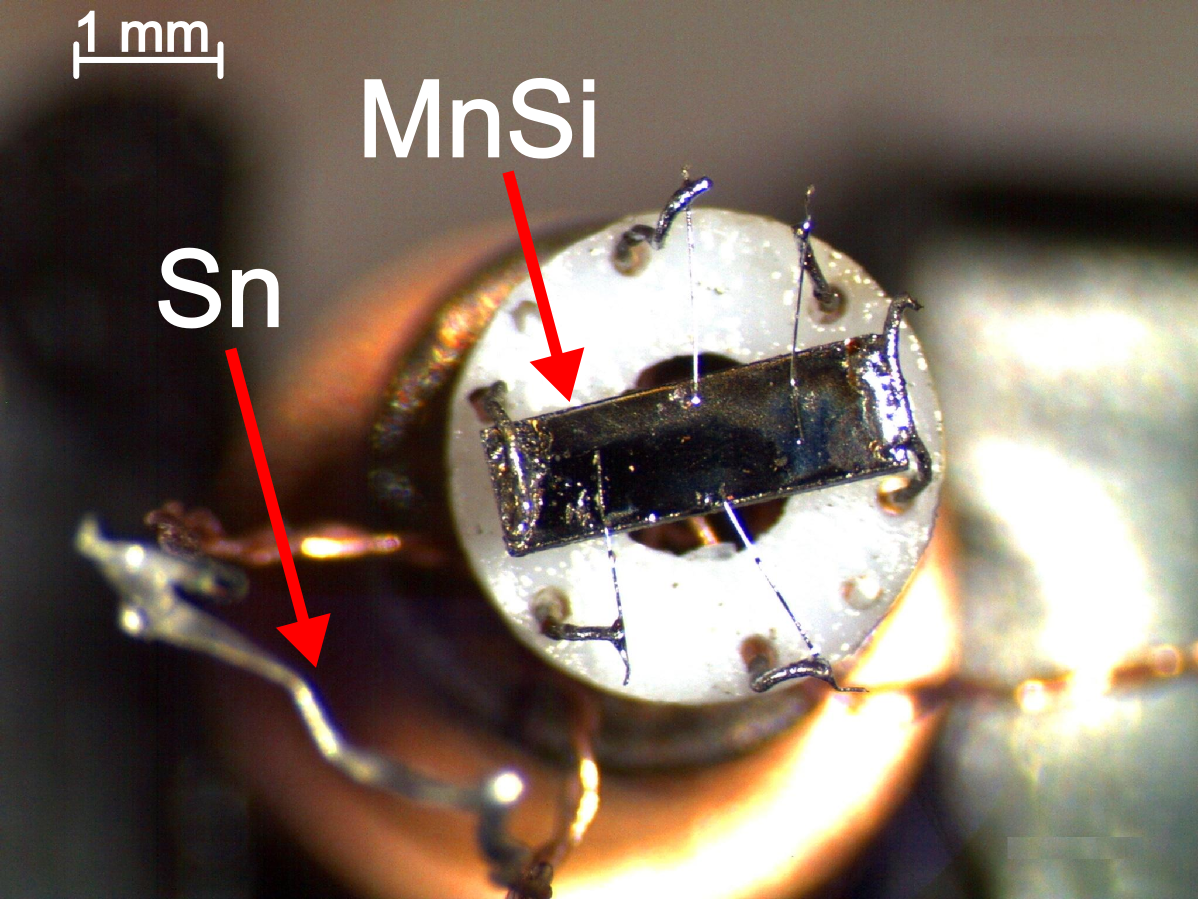}
\caption{(Color online) Photograph of a sample as mounted on the electrical feedthrough of the clamp-type pressure cell. A PTFE ring is used to fix the location of the electrical wires, which suspend the sample inside the pressure liquid. Also visible is the Sn sample used as pressure gauge, which was bent slightly to the side for better visibility.
}
\label{figure11}
\end{figure}

\section{Experimental Methods}
\label{methods}

The pressure dependence of the magnetotransport properties was studied down to 1.5\,K under magnetic fields up to 14\,T with an Oxford Instruments variable temperature insert (VTI) as combined with a superconducting magnet system. The temperature of the sample was monitored with a calibrated Cernox sensor, closely attached to the pressure cell. The magnetoresistance and the Hall voltage of the MnSi samples were measured simultaneously in a conventional six-terminal configuration using standard digital lock-in technology. Impedance matching low-noise signal transformer were used to increase the signal-to-noise ratio. Low excitation currents and frequencies were used to avoid resistive heating and parasitic signal pickup, respectively. 

Field sweeps were recorded for increasing and decreasing fields; temperature sweeps were recorded at positive and negative field values. This permitted to subtract longitudinal voltage components of the transverse contacts due to unavoidable contact misalignment by antisymmetrizing the transverse voltage pickup. Vice versa, transverse signal components at the longitudinal voltage contacts were corrected by symmetrizing the signal. We note that the result of the antisymmetrization/symmetrization procedure, by design, is a single field dependence from negative to positive field. The sign of the Hall signal was checked very carefully \cite{clarification}. In the following, the antisymmetric transverse signal is referred to as {\rxy} and the symmetric longitudinal signal is referred to as {\rxx}.

\begin{figure}
\includegraphics[width=0.35\textwidth]{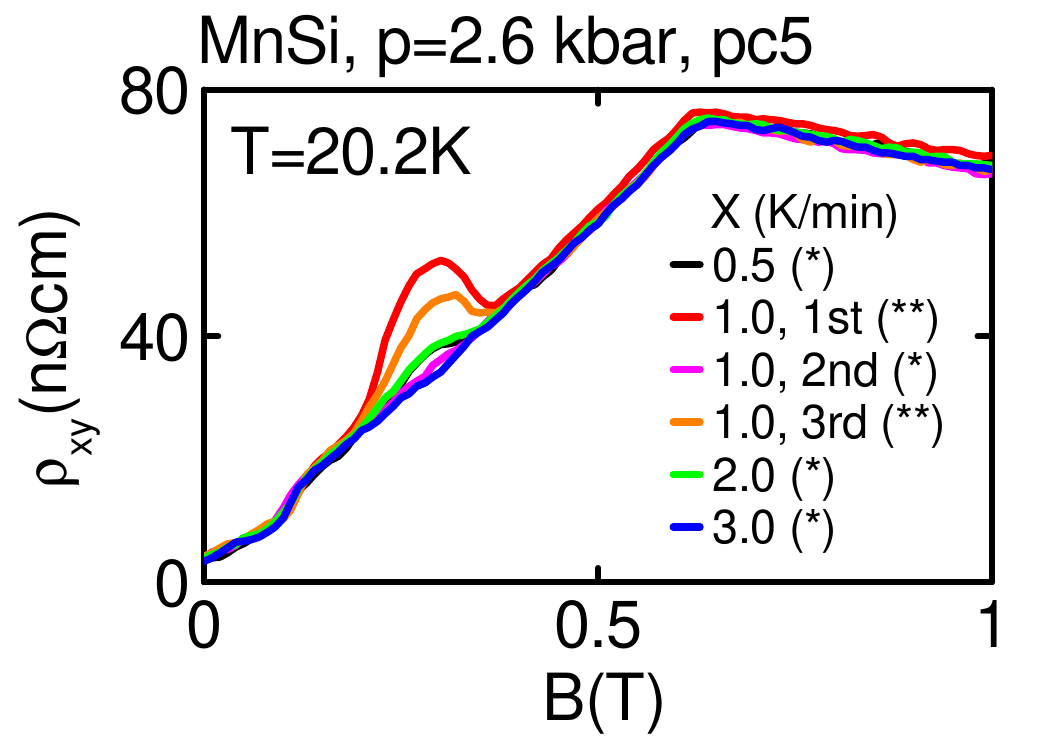}
\caption{(Color online) Typical Hall signal of a low-RRR MnSi sample in the regime of the A phase (i.e., the skyrmion lattice phase) as recorded following different cooling procedures of the pressure cell. For further details, see main text. All data reported in this paper were recorded after cooling the pressure cell through the freezing point of the methanol/ethanol mixture at 1\,K/min (red line).
}
\label{figure0}
\end{figure}

The single crystals studied were grown by optical float-zoning under ultrahigh vacuum (UHV) compatible conditions \cite{Neubauer:RSI2011}. Typical residual resistivity ratios (RRR) of our samples were in the range 40 to 300. The origin of the different RRRs can thereby be associated with the precise starting composition before float-zoning as determined in a careful systematic study to be reported elsewhere \cite{Bauer:PhD}. The differences we observe in the magnetotransport properties may be grouped into low-quality MnSi with low RRRs around $\sim 45$ and high-quality MnSi with RRRs above $\sim 90$. The latter allow to connect the topological Hall effect under pressure with the skyrmion lattice phase (the A phase) at ambient pressure as described in the main part of our paper.

In our transport measurements, platelet-shaped samples were studied with typical dimensions 2.8\,mm long, 1\,mm wide and less than 0.2\,mm thick. They were oriented such that the magnetic field was applied perpendicular to the platelet and parallel to {\ooz}. We have chosen this direction since neutron scattering under pressure shows that the easy axis of MnSi remains unchanged along {\ooo} up to $\sim$12\,kbar \cite{Fak:JPCM2005,Pfleiderer:Nature2004}, the {\ooz} axis is neither a magnetically hard nor soft axis for the pressure range studied, i.e., the crystallographic orientation is not distinct in any way. Electric currents were applied either along a {\ozz} or {\ooz}. The effects of demagnetizing fields were corrected by approximating the sample shape with a rectangular prism \cite{Aharoni:JAP98}. 

Conventional nonmagnetic Cu:Be clamp cells were used to study the pressure dependence of the magnetotransport properties. For the transport measurements, eight pressure cells were assembled. A detailed list of the pressure cells and pressures is given in Table\,\ref{table:samples}, where only pressures relevant to the work reported here are listed. The applied pressure was inferred from the combination of the superconducting transition of Sn measured resistively as well as a comparison of the helimagnetic transition temperature of MnSi with previous studies. For all pressure points investigated, we recorded at first the resistivity as a function of temperature at $B=0$ (data is not shown to safe space and because they are perfectly consistent with all previous studies). The transition temperature {\tc} referred to in the following was thereby determined from {\rxx} consistent with all previous studies. Further details of the temperature dependence will be addressed in Sec.\,\ref{results}.

\begin{table*}
\centering
\caption{Pressure cells prepared for our magnetotransport and magnetization measurements under pressure. Pressures are stated in the order in which they were applied. The residual resistivity ratio (RRR) was determined at the lowest pressure as the ratio $\rho_{\rm xx} \left( T=280\,\rm{K} \right) / \rho_{\rm xx} \left( T \rightarrow 0 \right)$. FI: Fluorinert FC72:FC84 mixture at a 1:1 volume ratio. ME: methanol:ethanol mixture at a 4:1 volume ratio. The current was applied along the long direction of the sample, the field was applied along the shortest direction.
\label{table:samples}}
\begin{tabular}{lccccc}
\hline\noalign{\smallskip}
\hline\noalign{\smallskip}
Pressure & & Sample size (${\rm mm^3}$)& & Pressure & \\
cell (pc) & RRR & $l\times w\times t$& Orientation & medium & Pressures (kbar)\\
\hline\noalign{\smallskip}
1 & $\approx$93 & $2.9 \times 1.0 \times 0.22$ & $B \parallel t\parallel \langle 110 \rangle$, $I \parallel l\parallel \langle 100 \rangle$ & FI & 6.6 \\
2 & $\approx$92 & $2.8 \times 1.0 \times 0.25$ & $B \parallel t\parallel \langle 110 \rangle$, $I \parallel l\parallel \langle 100 \rangle$ & FI & 7.0, 10.0 \\
3 & $\approx$300 & $2.5 \times 1.0 \times 0.20$ & $B \parallel t\parallel \langle 110 \rangle$, $I \parallel l\parallel \langle 100 \rangle$ & ME & 10.7 \\
4 & $\approx$300 & $2.7 \times 0.9 \times 0.20$ & $B \parallel t\parallel \langle 110 \rangle$, $I \parallel l\parallel \langle 100 \rangle$ & ME & 8.1 \\
5 & $\approx$45 & $2.8 \times 1.0 \times 0.20$ & $B \parallel t\parallel \langle 110 \rangle$, $I \parallel l\parallel \langle 110 \rangle$ & ME & 7.6, 6.7, 5.9, 5.7, 5.1, 4.6, 3.4, 2.9, 2.6, 2.2, 0.3 \\
6 & $\approx$40 & $2.7 \times 1.0 \times 0.20$ & $B \parallel t\parallel \langle 110 \rangle$, $I \parallel l\parallel \langle 100 \rangle$ & ME & 5.2, 7.4, 9.6, 11.0, 11.2, 12.8, 0.4 \\
7 & $\approx$45 & $2.8 \times 1.0 \times 0.20$ & $B \parallel t\parallel \langle 110 \rangle$, $I \parallel l\parallel \langle 110 \rangle$ & ME & No pressures below $p^*$ \\
8 & $\approx$150 & $2.9 \times 1.1 \times 0.20$ & $B \parallel t\parallel \langle 110 \rangle$, $I \parallel l\parallel \langle 100 \rangle$ & ME & 11.5, 10.3, 9.3, 7.8, 6.1, 4.9, 3.7, 0.5\\
M & $\approx$70 & $6.0 \times 1.0 \times 1.0$ & $B \parallel t\parallel \langle 100 \rangle$ & ME & 0.0, 4.05, 7.50, 10.13, 11.80\\
\noalign{\smallskip}\hline
\noalign{\smallskip}\hline
\end{tabular}
\end{table*}

For our high-pressure measurements, the platelet-shaped samples were mounted perpendicular to the cylinder axis of the pressure cell. Shown in Fig.\,\ref{figure11} is a typical setup as seen from the top of the electrical feedthrough (along the cylinder axis of the pressure cell). The current carrying Cu leads (diameter 0.120\,mm) were soldered directly to the small faces of the sample. Pt wires (diameter 0.025\,mm ) were spot welded to the sample providing tiny, nonsuperconducting voltage contacts. The Pt wires were in turn soldered to Cu leads exiting the pressure cell. A polytetrafluoroethylene (PTFE) disk was used to guide the electrical leads exiting the feedthrough inside the pressure cell, such that the sample was stabilised against accidental tilting during pressure changes. As the main advantage of this setup, the sample was essentially floating freely in the pressure transmitter suspended by the current leads only. This minimizes parasitic effects anticipated of differences of compressibilities when glueing the sample directly to a  support structure [empirically, the latter is long known to drive changes of the magnetic easy axis in MnSi (see, e.g., Ref.\,\cite{Fak:JPCM2005})]. Also visible in this picture is the Sn sample used to determine the pressure (for better visibility  it was slightly bent to the side).

The geometry factors used for calculating {\rxx} and {\rxy} were at first determined with a light microscope. To account for small systematic differences between samples and to permit better comparison of data recorded for different samples, we adjusted the longitudinal resistivity at 35\,K to the pressure dependence determined in Ref.\,\cite{Pfleiderer:PRB97}, $\rho_{xx}({B=0},\,\,35\,{\rm K},\,\,p)=48\,{\rm \mu\Omega\,cm}-0.8\,{\rm \mu\Omega\,cm\,kbar^{-1}}\, p$. Likewise, we adjusted the geometry factor used for the Hall data by virtue of a comparison with data recorded with two pressure cells over the full pressure range at a field of 13.5\,T and a temperature of 2.8\,K, notably $\rho_{xy}({\rm 13.5\,T,\,\,2.8\,K},\,\,p)=0.19\,{\rm \mu\Omega\,cm}-0.002\,{\rm \mu\Omega\,cm\,kbar^{-1}}\, p$. The adjustments were no larger then 10\,\%.

To ensure ideal pressure conditions, the majority of our experiments were  carried out with a 4:1 methanol:ethanol mixture as pressure transmitter, reported to provide the best pressure homogeneity in the pressure range of interest as compared with other organic liquids \cite{Thomasson}. Data recorded with the methanol:ethanol pressure transmitter is labelled ME. In addition, we performed a few experiments with a 1:1 mixture of Fluorinert FC72 and FC84 as pressure transmitter, denoted by FI. This mixture has been reported to provide, in principle, fairly uniform pressure conditions up to 15\,kbar \cite{Sidorov:JPCM2005}. However, as summarized in Fig.\,\ref{figure8}, we find discrepancies consistent with local strains in the Fluorinert pressure transmitter as compared with data recorded with methanol:ethanol. In fact, it is important to note that a single-component Fluorinert (FC77) was used in previous magnetotransport studies in MnSi under pressure \cite{Lee:PRL09}, which is known to provide even less uniform pressure conditions. 

In our studies we found that the topological Hall signal varied sensitively with the cooling procedure. Following careful tests, the largest topological Hall signal was systematically observed under two conditions: first, cooling the pressure cells from room temperature, and second, cooling at sufficiently slow cooling rates between 200 and 100\,K, which covers the solidification temperature of the methanol:ethanol mixture. In fact, this observation was insensitive to the pressure transmitters we tested.  We have therefore systematically recorded all data presented in this paper after initially cooling the sample from room temperature with a rate of 1\,K/min between 200 and 100\,K. We presume that this procedure minimizes local strains that arise otherwise from the solidification of the pressure transmitter. 

Typical data illustrating the sensitivity to the cooling procedure are shown in Fig.\,\ref{figure0}, where data recorded after cooling from room temperature are marked (**) and data recorded after heating to 250\,K followed by cooling from 250\,K are marked with (*). Data of one sequence shown in Fig.\,\ref{figure0} consisted in cooling the cell always at 1\,K/min, first from room temperature, second from 250\,K after heating the cell up from 2\,K, and third again from room temperature having cooled the cell to 2\,K again. Data of the other sequence shown in Fig.\,\ref{figure0} were recorded after the cell was heated to 250\,K, but the subsequent cooling was done at three different rates of 0.5, 2, and 3\,K/min, respectively. It is important to emphasize that the pressure cell never reached room temperature in the second sequence. Thus, the pressure transmitter must have retained some of the frozen-in pressure inhomogeneities, which completely vanish when heating the pressure cell consequently to room temperature.

The pressure dependence of the magnetization was, finally, measured with a non-magnetic Cu:Be miniature clamp cell \cite{Pfleiderer:JPCM2004} in an Oxford Instruments vibrating sample magnetometer (VSM) at temperatures down to 2.3\,K for magnetic fields up to 9\,T under pressures up to 12\,kbar. The empty pressure cells were measured and their signal (even though tiny) carefully subtracted to determine the signal of the MnSi sample. Typical magnetization samples had the shape of a bar ($6\times1\times1\,{\rm mm^3})$ oriented along the cylinder axis of the pressure cell and thus parallel to the applied magnetic field. 

\section{Experimental Results}
\label{results}
\subsection{Magnetic field dependence}

\begin{figure}
\includegraphics[width=0.4\textwidth]{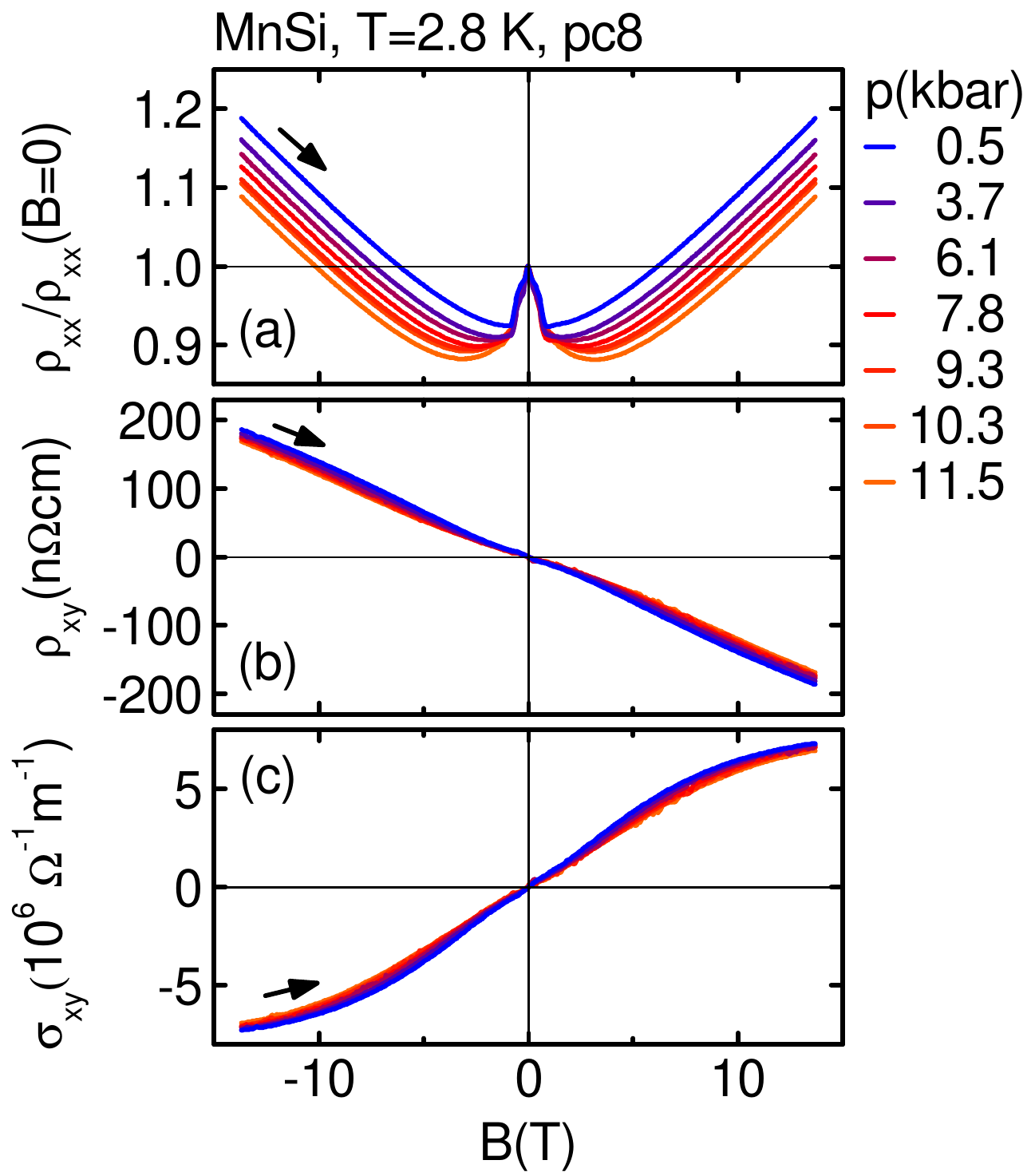}
\caption{(Color online) Typical magnetotransport properties of MnSi at 2.8\,K  under applied magnetic fields up to 14\,T for various pressures up to 11.5\,kbar. (a) Magnetoresistance {\rxx} as normalised to $B=0$. (b) Hall resistivity {\rxy} as a function of magnetic field. Only a very weak pressure dependence is observed. (c) Hall conductivity {\Sxy} calculated from the data shown in panels (a) and (b).
}
\label{figure1}
\end{figure}

We begin with the magnetotransport properties at a low temperature of 2.8\,K under magnetic fields up to 14\,T, the largest fields measured. As shown in Fig.\,\ref{figure1}\,(a), the magnetoresistance {\rxx} displays a maximum with respect to zero magnetic field, followed by a shallow minimum and an increase at high fields. With increasing pressure the same qualitative field dependence is observed up to $\sim12\,{\rm kbar}$, where the magnetoresistance up to 14\,T increases. 

\begin{figure}
\includegraphics[width=0.48\textwidth]{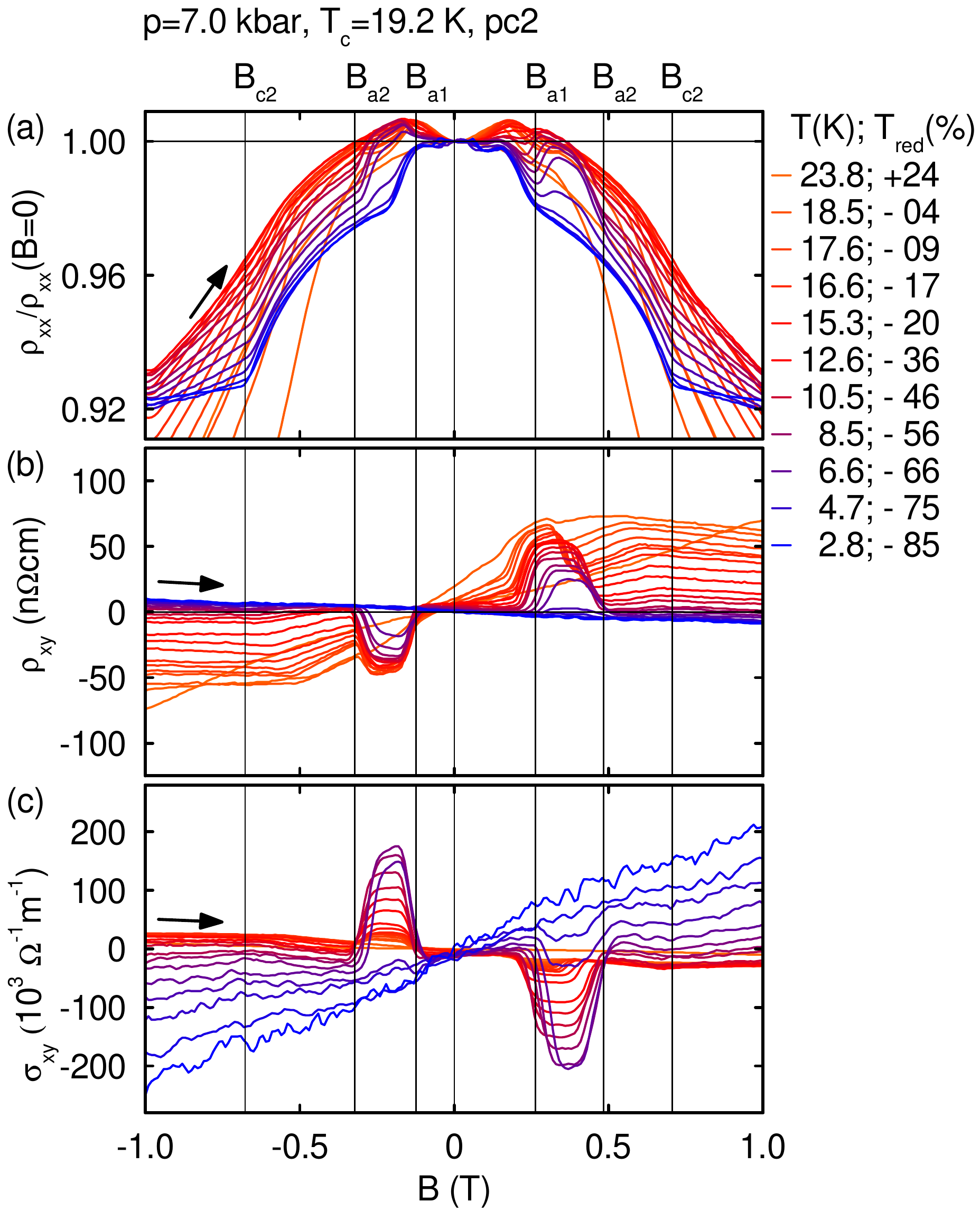}
\caption{(Color online) Typical magnetotransport data in single-crystal MnSi at a pressure of 7\,kbar under applied magnetic fields up to 1\,T for various temperatures. (a) Magnetoresistance {\rxx} as a function of field for various temperatures. (b) Hall resistivity {\rxy} as a function of field at various temperatures. Note the giant top-hatshaped topological contribution. (c) Hall conductivity {\Sxy} calculated from the data shown in panel (a) and (b).
}
\label{figure2}
\end{figure}

For the same conditions, the Hall resistivity {\rxy} decreases at 2.8\,K over the entire field range up to 14\,T as shown in Fig.\,\ref{figure1}\,(b). With increasing pressure {\rxy} displays a very weak pressure dependence. Most important, the high-field slope of {\rxy} and thus the effective charge carrier concentration in this field range are essentially unchanged [see also Fig.\,\ref{figure10}\,(b) below]. The Hall conductivity $\sigma_{\rm xy}=-\rho_{\rm xy}/(\rho_{\rm xx}^2+\rho_{\rm xy}^2)$ calculated from the magnetoresistance and Hall resistivity, shown in Fig.\,\ref{figure1}\,(c), is essentially featureless. A small nonlinear contribution at high fields provides evidence of an anomalous Hall contribution due to the uniform magnetization. As shown below, the anomalous Hall contribution vanishes with decreasing temperature and is therefore already very small at 2.8\,K. Consistent with the magnetization shown below, the Hall conductivity decreases weakly with increasing pressure.

We now turn to the detailed behavior in the vicinity of $T_c$ for low magnetic fields. Shown in Fig.\,\ref{figure2} are typical magnetotransport data for $p=7\,{\rm kbar}$ as a function of magnetic field up to 1\,T [temperatures are also stated as reduced values $T_{\rm red}=(T-T_{\rm c})/T_{\rm c}$]. At high temperatures, the transverse magnetoresistance {\rxx}, shown in Fig.\,\ref{figure2}\,(a), decreases with increasing magnetic field. For $T<T_{\rm c}$, the magnetoresistance increases at first gently up to {\bco}, levels off, and displays a shallow maximum in a field and temperature range somewhat larger than the skyrmion lattice phase at ambient pressure before decreasing further. At the lowest temperatures, the magnetoresistance decreases on the field scale shown here with distinct changes of slope at {\bco} and {\bct} (for clarity {\bco} is not marked in Fig.\,\ref{figure2}).

\begin{figure}
\includegraphics[width=0.45\textwidth]{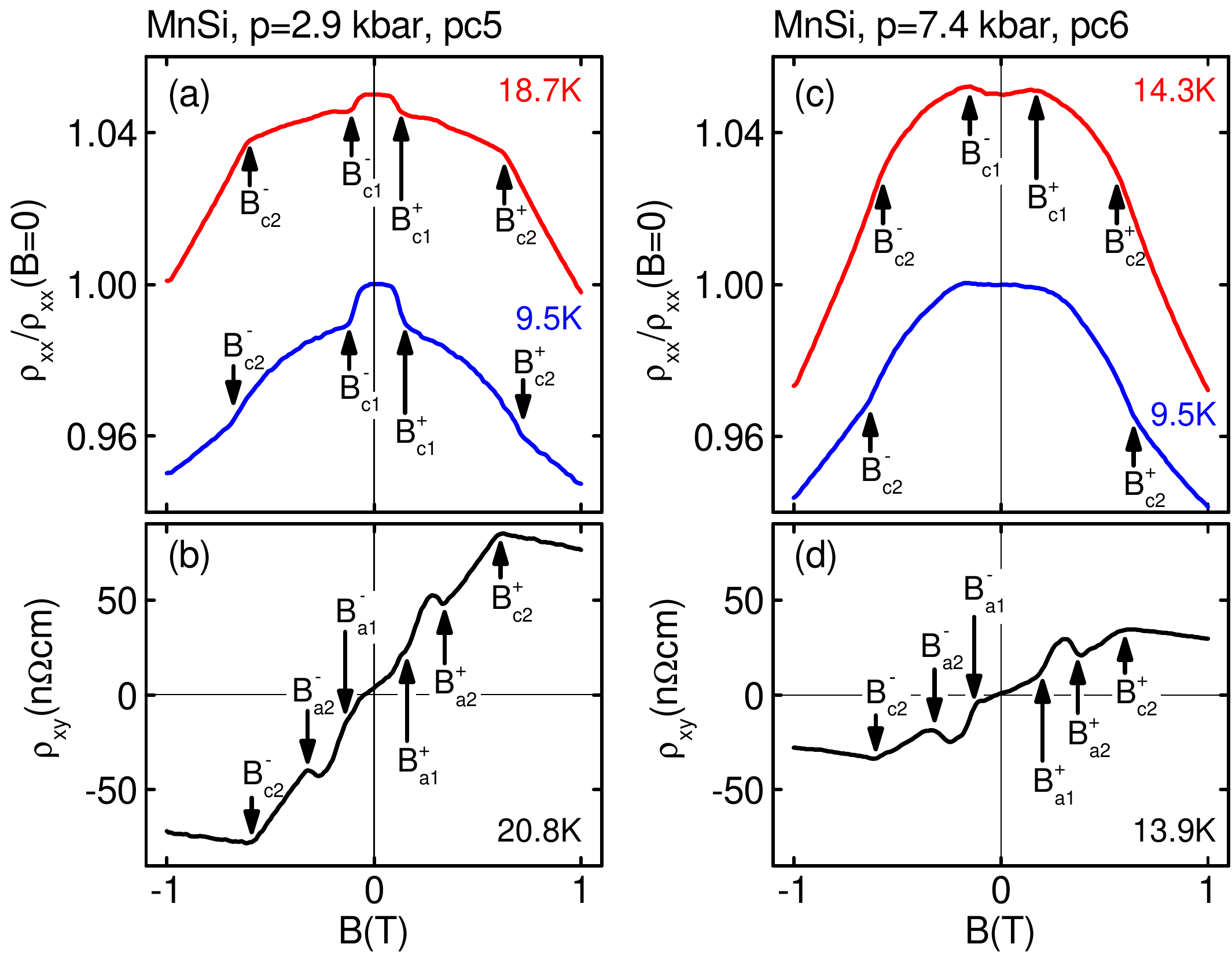}
\caption{(Color online) Typical magnetoresistance and Hall effect for two different pressures. The figure illustrates the definitions of the characteristic transition fields. (a) Magnetoresistance at 2.9\,kbar and two different temperatures. The magnetoresistance initially decreases and changes slope at {\bco}. (b) Hall resistivity at 2.9\,kbar. (c)  Magnetoresistance at 7.4\,kbar and two different temperatures. At the transition from the helical to the conical phase, a shallow maximum is observed. (d) Hall resistivity at 7.4\,kbar and 13.9\,K. Note that the upper curves in panels (a) and (c) have been shifted up for clarity.
}
\label{figure13}
\end{figure}

The Hall resistivity {\rxy}, shown in Fig.\,\ref{figure2}\,(b), displays a gradual field dependence with a pronounced  top-hat-shaped enhancement in a small field and temperature range {\bao} and {\bat}, somewhat larger than the skyrmion lattice phase at ambient pressure. In other words, with increasing field the enhancement appears abruptly at a field {\bao}, and vanishes again equally abruptly at a field {\bat}. The magnitude of the top-hat-shaped signal contribution is substantially larger than a similar signal contribution in the skyrmion lattice phase at ambient pressure \cite{Neubauer:PRL2009}. The signal size corresponds thereby quantitatively to the data reported in Ref.\,\cite{Lee:PRL09}. However, depending on the precise experimental conditions, the field range in which we observe the top-hat signal is smaller to that reported in Ref.\,\cite{Lee:PRL09}, where it existed all the way from {\bco} to {\bct} (we return to the importance of sample quality and pressure homogeneity for this effect below). 

To elucidate the origin of the large magnitude of the top-hat-shaped signal contribution, we show in Fig.\,\ref{figure2}\,(c) the Hall conductivity  $\sigma_{\rm xy}=-\rho_{\rm xy} /(\rho_{\rm xy}+\rho_{\rm xx})^{2}\approx -\rho_{\rm xy} /\rho_{\rm xx}^{2}$. The top-hat-shaped contribution  in {\Sxy} grows much stronger for lower temperature (and therefore lower $\rho_{\rm xx}$) than the signal in $\rho_{\rm xy}$. As discussed in the Introduction, for the intrinsic anomalous Hall effect, one expects a universal Hall signal in $\sigma_{\rm xy}$ independent of the scattering time $\tau$, while for the topological Hall effect $\rho_{\rm xy}$ is independent of $\tau$ (such that $\sigma_{\rm xy}$ increases proportional to $1/\rho_{\rm xx}^2$). Therefore, these data suggest that the top-hat signal can be identified with the topological Hall signal which is switched on and switched off when the system enters and leaves the skyrmion phase, respectively.

Shown in Fig.\,\ref{figure13} are typical data to illustrate the definition of the transition fields {\bco}, {\bct}, {\bao}, and {\bat}. Values for increasing and decreasing field strength are denoted by the superscripts ``+'' and ``-'', respectively. For the transition fields {\bco} and {\bct}, no hysteresis is observed at all pressures and temperatures studied. Moreover, the same qualitative field dependences are observed for all pressures studied up to $\sim$12\,kbar, with the exception of {\rxx} below {\bco}. This is illustrated in Fig.\,\ref{figure13}\,(a) and (c), where the magnetoresistance drops slightly at {\bco} at low pressures, while it displays a shallow minimum up to {\bco} for higher pressures, respectively. 

\begin{figure}
\includegraphics[width=0.3\textwidth]{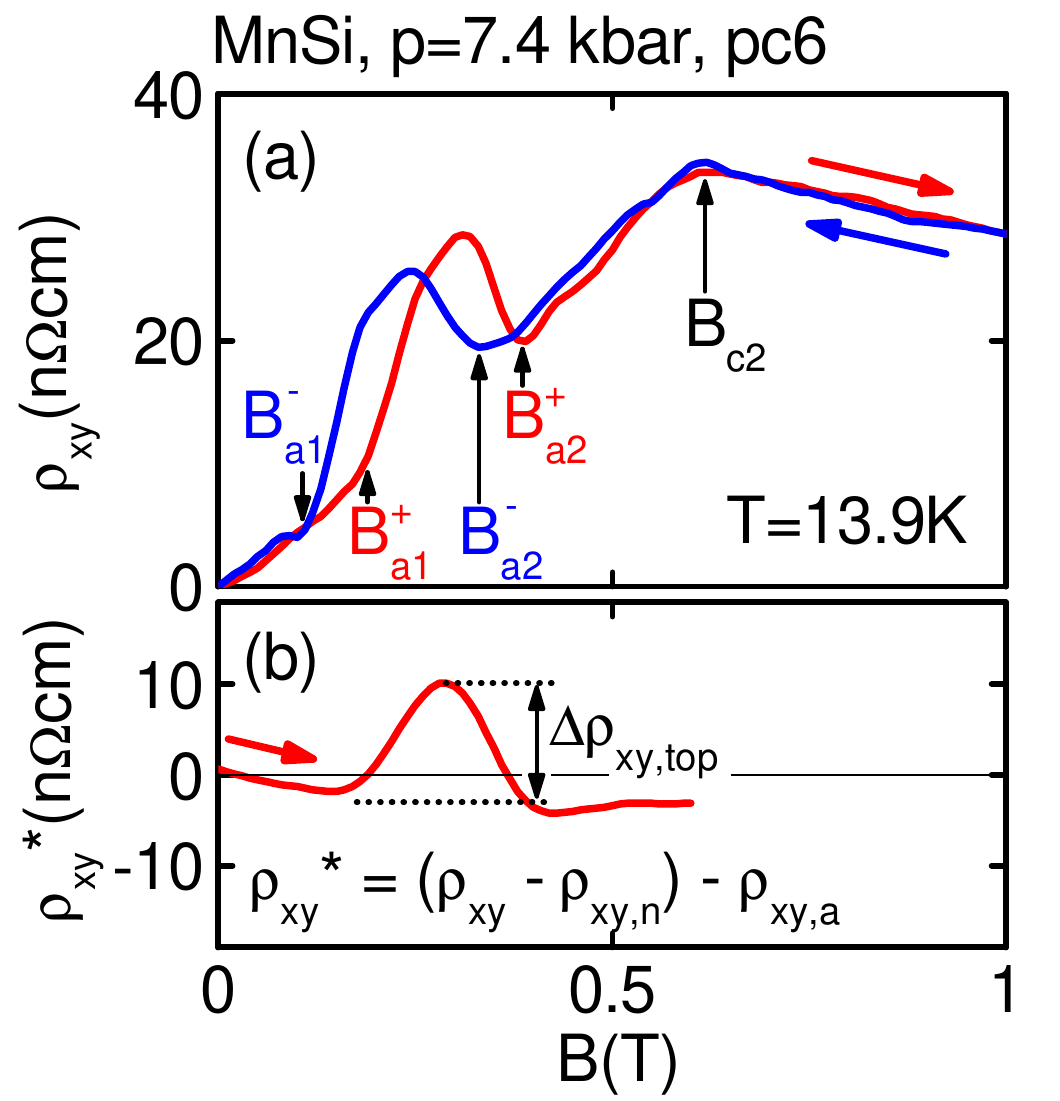}
\caption{(Color online) Typical field dependence of Hall data in the regime of the A phase. Shown are data for increasing field (red arrow) and decreasing field (blue arrow). (a) In the regime of the A phase, a clear hysteresis is observed at {\bao} and {\bat}, while {\bct} and the data outside the A phase are not hysteretic. (b) The size of \rxytop was determined as the peak height of {\rxy} after subtracting the estimated normal and anomalous Hall contributions.
}
\label{figure12}
\end{figure}

Closer inspection of the top-hat-shaped topological signal reveals the presence of hysteresis at {\bao} and  {\bat} as shown in Fig.\,\ref{figure12}. This suggests that the transitions at {\bao} and {\bat} are first order (cf. Ref.\,\cite{Bauer:PRB12}). In comparison, the features of {\bco} and {\bct} are not hysteretic (the transitions may nevertheless be very weakly first order). In passing, we note that high-pressure studies reported previously \cite{Lee:PRL09} have not addressed the question of hysteresis at all.

\begin{figure}
\includegraphics[width=0.47\textwidth]{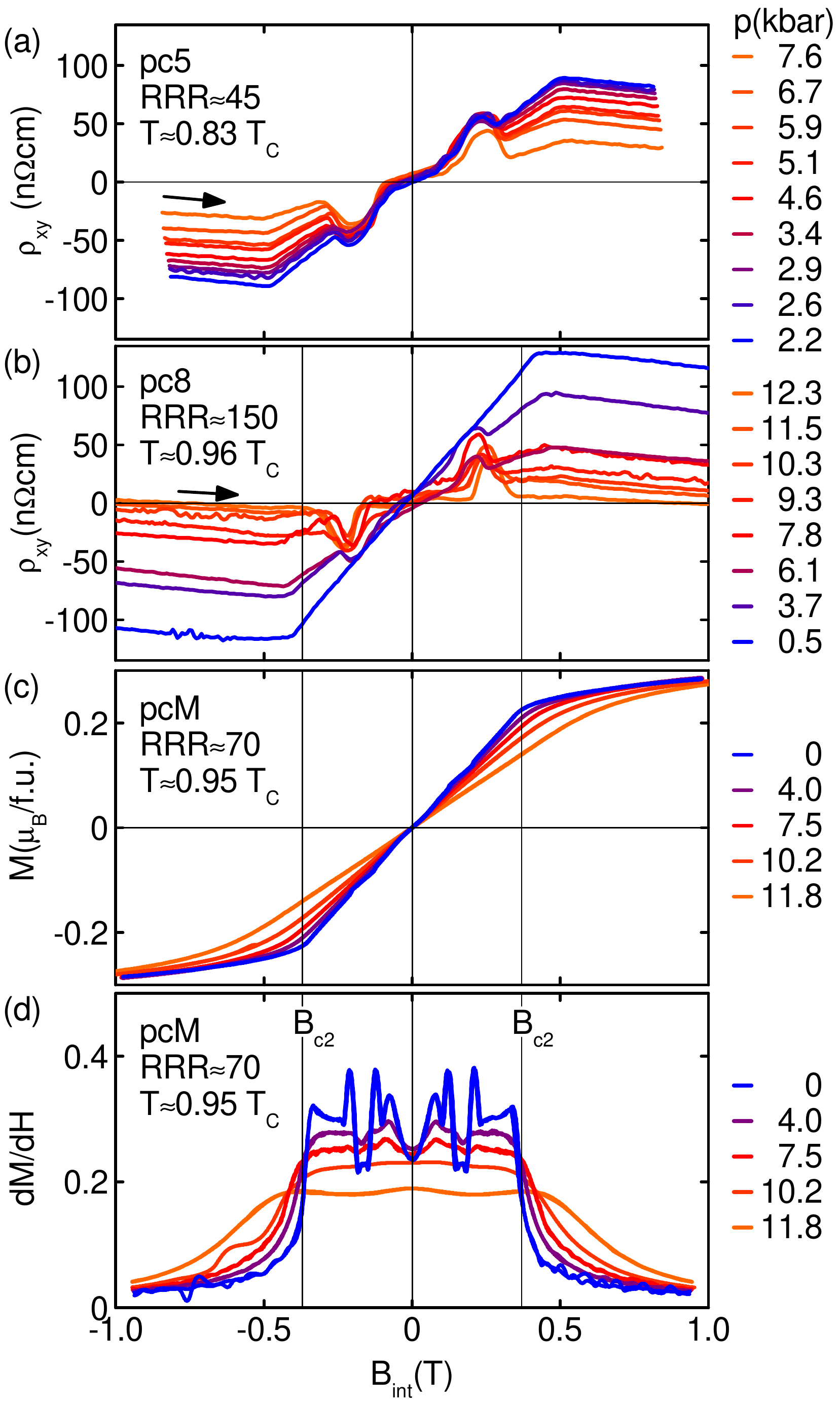}
\caption{(Color online) Typical Hall resistivity and magnetization data at similar reduced temperatures below {\tc} for various pressures. The Hall data are shown for increasing field strength; the magnetization data is shown for both sweep directions. Note the data are shown as function of estimated internal field, i.e., demagnetizing fields were corrected. (a) Hall resistivity for a low-quality sample with RRR=45. The top-hat-shaped contribution extends over a wide range much wider than for the A phase at ambient pressure. (b) Hall resistivity for a sample with RRR=150.  (c) magnetization of sample with RRR=70. (d) Susceptibility calculated from the data shown in panel (c). The pronounced narrow maxima at the boundary of the skyrmion lattice phase vanish with increasing pressure, while the field value of the transition remains unchanged.}
\label{figure3}
\end{figure}

It is now instructive to consider changes of the magnetic field dependence of {\rxx} and {\rxy} with changes of pressure as summarized in Fig.\,\ref{figure3}. We thereby focus at first on the top-hat contribution and return to the rest of the Hall signal further in the following. For the pressure range of our study, the top-hat-shaped signal contribution becomes maximal at roughly the same reduced temperature below {\tc} and vanishes in the field-dependent data when decreasing the temperature further. It is thereby important to note that the reduced temperature of the maximum top-hat Hall signal depends on the sample quality. Namely, for a sample with low RRR ($\sim45$) akin that studied in Ref.\,\cite{Lee:PRL09}, the maximum top-hat contribution is located roughly $\sim17\,\%$ below $T_c$. The associated field dependence at 17\,\% below $T_c$ for various pressures is shown in Fig.\,\ref{figure3}\,(a). With increasing pressure, the size of the top-hat contribution increases, where we return to the detailed pressure dependence below. At the same time the field range of the top-hat signal contribution is rather wide for this low-quality sample.

In contrast, samples with much higher RRRs ($\sim150$) display the maximum top-hat contribution about $\sim$4\,\% below $T_c$. Typical field dependencies for a temperature around 4\,\% below $T_c$ are shown in Fig.\,\ref{figure3}\,(b). With increasing pressure, the top-hat contribution increases. Here, the field range of the top-hat contribution is smaller than for the lower-quality sample. Taken together, the temperature and field range of the top-hat contribution of the high-RRR samples are much closer to the field and temperature range of the skyrmion lattice phase at ambient pressure than for low-RRR samples. This clearly demonstrates a high sensitivity of the top-hat contribution to the sample quality. 

In view of the importance of the sample quality we have performed preliminary tests of the role of the pressure transmitter. Namely, we used a Fluorinert mixture (denoted FI in the figures), which is known to provide inhomogeneous pressures, for some measurements. Here, the field range of the top-hat contribution is even wider as compared with the ME mixture (cf. Fig.\,\ref{figure2}). In comparison, data reported in Ref.\,\cite{Lee:PRL09}, where the top-hat-shaped topological contribution was observed all the way up to {\bct}, were recorded with a single-component Fluorinert pressure transmitter. 

We finally turn to the Hall signal outside the field range of the top-hat contribution (i.e., outside of the skyrmion phase). This part of the Hall signal varies strongly with pressure for the high-quality sample, while it changes only weakly for the low-quality sample as shown in Figs.\,\ref{figure3}\,(a) and (b). Since the data shown in these figures were recorded at the same reduced temperature below {\tc}, and {\tc} decreases by a factor of 2 between $p=0$ and 12\,kbar, the decrease of the Hall signal outside the field range of the top-hat contribution is essentially a consequence of the decrease of {\rxx} with decreasing {\tc}. Thus, the non-top-hat part of the Hall signal is characteristic of a normal plus an (intrinsic) anomalous Hall signal. 

The difference between the topological and anomalous contribution to the Hall signal is strongly supported by the magnetization as a function of magnetic field, shown in Fig.\,\ref{figure3}\,(c) for various pressures at a temperature 5\,\% below {\tc}. With increasing pressure and thus decreasing {\tc}, the magnetization is slightly reduced, while the magnetic field dependence changes very little qualitatively.  The non-top-hat part of the Hall signal therefore qualitatively tracks the magnetization as expected of a dominant anomalous Hall contribution. Unfortunately, it is not possible to carry out a full analysis, in which the anomalous Hall contribution is directly calculated from the magnetization, since the pressures and demagnetizing fields differ between the magnetization data and the magnetotransport data.

More subtle changes of the magnetization with increasing pressure may be revealed by the susceptibility {\dmdb}, calculated from the magnetization as shown in Fig.\,\ref{figure3}\,(d). As recently established in a comprehensive study, {\dmdb} provides a reliable probe of phase boundaries as compared with the ac susceptibility \cite{Bauer:PRB12}. Namely, at ambient pressure, {\dmdb} displays sharp spikes at the boundary of the skyrmion lattice phase, characteristic of a first-order transition. With increasing pressure, these spikes smear out and vanish while the transition fields remain essentially unchanged [this corresponds also to the smearing reported recently in the resistivity at $T_c$ (Ref.\,\cite{Petrova:PRB2012})]. The simultaneous presence of the large top-hat signal  suggests that the spikes in {\dmdb} vanish due to small pressure inhomogeneities which do not affect the main conclusions of the study reported here. The magnetization, hence, suggests that the top-hat Hall contribution is not the result of a possible contribution in the uniform magnetization. In fact, in the field range of the large topological Hall signal, the magnetization decreases $\sim20\,\%$ with increasing pressure. This suggests strongly that the large topological Hall signal is not connected with changes of the local spin polarization. 

The size of the top-hat signal contribution of {\rxy} may be estimated by subtracting the normal and anomalous Hall signal in two steps as illustrated in Fig.\,\ref{figure12}. The normal Hall contribution was first inferred from {\rxy} at high fields and a linear field dependence subtracted. For the resulting signal, a linear field dependence was assumed up to {\bct} and subtracted. The resulting signal is dominated by the top-hat contribution. The size of this signal, $\Delta\rho_{\rm xy}^{\rm top}$, is estimated as shown in Fig.\,\ref{figure12}\,(b). We return to the pressure dependence of  $\Delta\rho_{\rm xy}^{\rm top}$  below.

Taken together, we observe a large top-hat (topological) Hall signal even in high-quality single crystals under (essentially) homogeneous pressure conditions. We are thereby empirically able to attribute the extended field region where a topological Hall signal was found in previous studies under pressure, and which appeared to be inconsistent with the field region of the topological Hall signal at ambient pressure, to a combination of sample quality and anisotropic pressure conditions. The role of sample quality, discussed in more detail in the next section, suggests that the pinning due to disorder and/or pressure inhomogeneities and local uniaxial strain arising due to local difference in the compressibility strongly affect the field and temperature range where the  top-hat Hall contribution is observed. While  the (meta)stability of the corresponding phase is strongly affected, the magnitude of this signal is rather insensitive to sample quality (a more detailed discussion of the signal size will be presented in the following). 


\subsection{Temperature dependence}

It is now instructive to explore the temperature dependence of the top-hat contribution of the Hall signal in further detail. Two measurement protocols have thereby been used, as the Hall signal is sensitive to the field and temperature history. In the first protocol, denoted zfc/fh, the sample was first zero-field cooled, the magnetic field applied next at the lowest temperature accessible (typically 2\,K) and data recorded while heating the sample in the applied field. In the second protocol, denoted fc/fh, the sample was cooled down in the applied field and data recorded while heating the sample in the same unchanged applied field. This way, data were recorded while heating in the same way, minimizing systematic errors between zfc/fh and fc/fh. In order to justify this approach, we have confirmed that data recorded during field cooling agree with data recorded while field heating after field cooling. 

\begin{figure}
\includegraphics[width=0.27\textwidth]{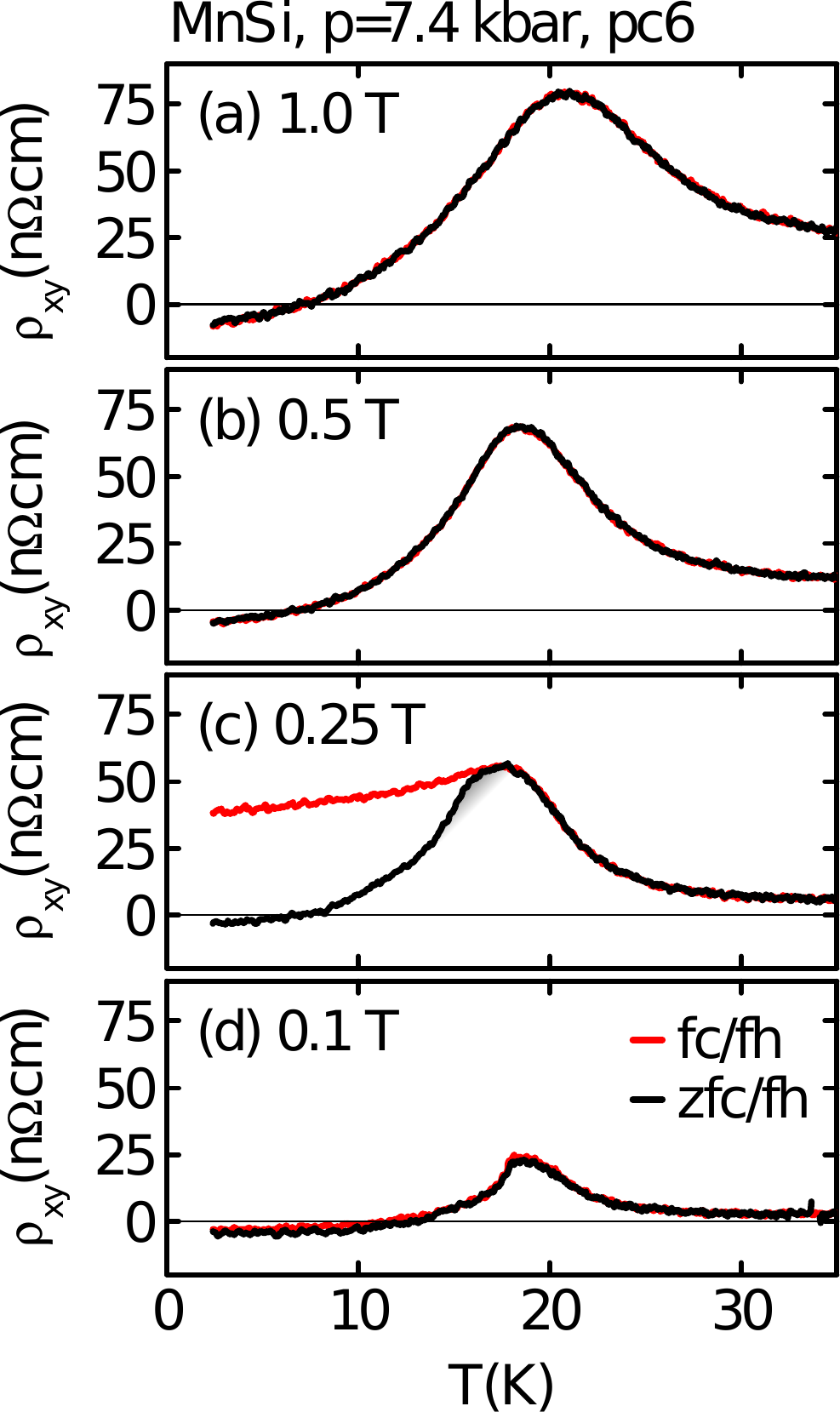}
\caption{(Color online) Typical temperature dependence of the Hall signal as recorded while heating in an applied field after field cooling or zero-field cooling, denoted as fc/fh and zfc/fh, respectively. No difference is observed at very small and sufficiently high fields as shown in panels (a), (b), and (d). In the field range of the A phase, the Hall signal remains high and essentially unchanged down to the lowest temperatures under field cooling [panel (c)].
}
\label{figure5}
\end{figure}

\begin{figure}
\includegraphics[width=0.45\textwidth]{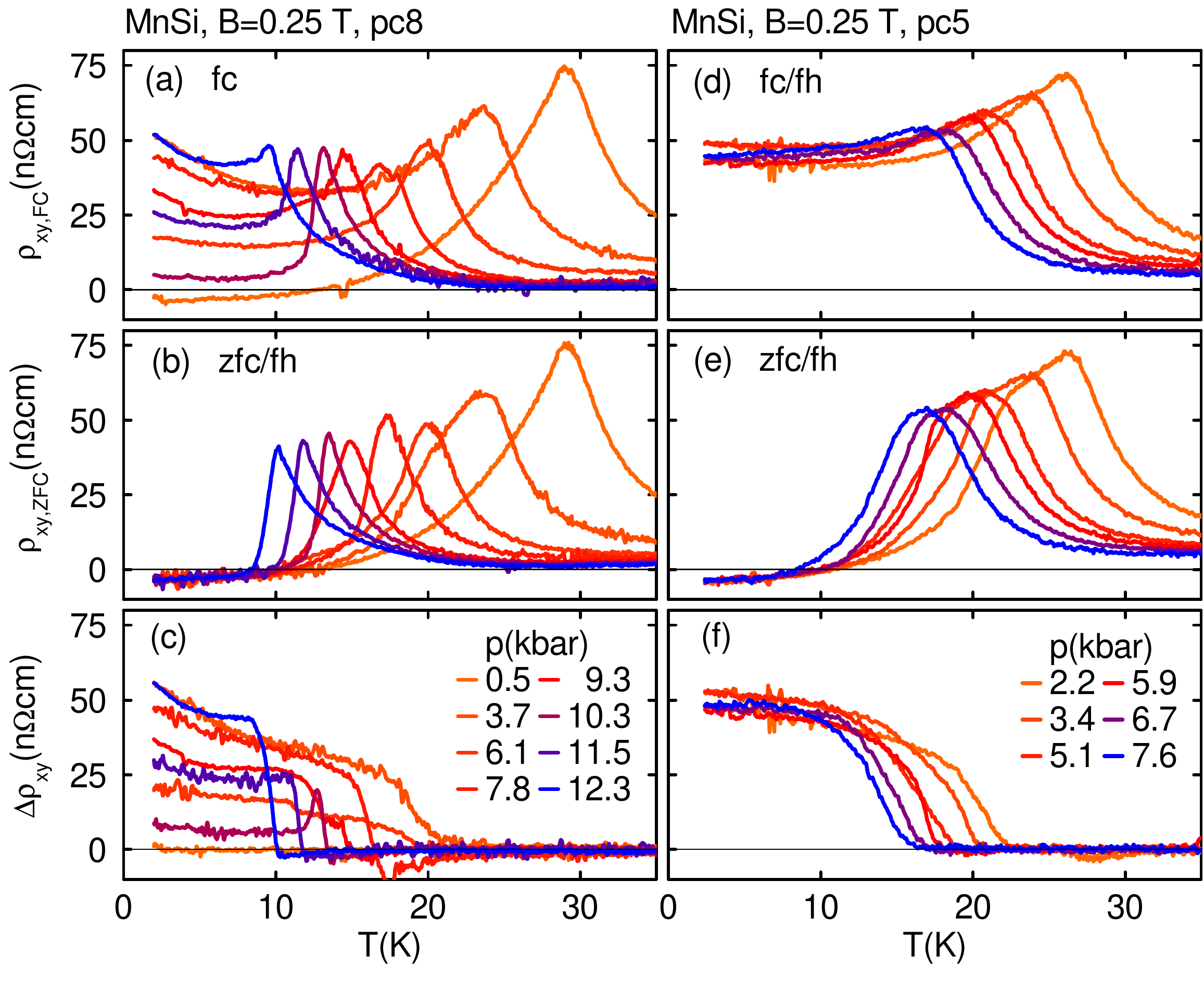}
\caption{(Color online) Hall resistivity as a function temperature for different sample qualities and various pressures. (a) Hall resistivity after field cooling (fc) for a sample with RRR=150. (b) Hall resistivity under field heating after zero-field cooling (zfc) for a sample with RRR=150. (c) Difference of panels (a) and (b). Pressures of panels (a)-(c) are indicated in panel (c). (d)-(f) are same as panels (a)-(c) for a sample with RRR=45. Pressures of panels (d)-(f) are indicated in panel (f).
}
\label{figure4}
\end{figure}

Typical temperature dependences are shown in Fig.\,\ref{figure5}. For all magnetic fields, the Hall signal is essentially dominated by a broad maximum in the vicinity of {\tc}. No difference is observed between zfc/fh and fc/fh data for magnetic fields outside the field range in which the top-hat contribution to the Hall signal is seen in field sweeps just below $T_c$ [Fig.\,\ref{figure5}\,(a), (b), and (d)]. However, for magnetic fields in the range of the top-hat contribution, the Hall signal for fc/fh retains a large value below the broad maximum, while the data recorded under zfc/fh decreases below the broad maximum. The difference suggests that the top-hat Hall signal survives under field cooling as a metastable state down to the lowest temperatures. This is consistent with the picture that once the sample has been prepared in the skyrmion lattice phase, thermal fluctuations are not successful to unwind the magnetic structure, which therefore remains as a metastable state (see Sec.\,\ref{phasediagram} below).

Trying various other combinations of field and temperature histories, we find no other possibility to prepare a similarly large Hall signal at the lowest temperatures as compared with field cooling. Interestingly, however, similar metastable behavior has been observed in small-angle neutron scattering studies of the skyrmion lattice phase in {\fcs} \cite{Muenzer:PRB2010}. 

Typical data illustrating the metastable temperature dependence at various pressures are shown in Fig.\,\ref{figure4} for two different sample qualities at a field of 0.25\,T. For pressures exceeding several kbar, the metastable behavior emerges. When field cooling (fc) at a slow rate in the field range of the skyrmion lattice phase, {\rxy} increases for decreasing temperature with a maximum just above {\tc}, typically retaining a high value down to the lowest temperatures measured [Figs.\,\ref{figure4}\,(a) and (d)]. In contrast, data recorded under slow field heating after zero-field cooling (zfc/fh) drops to a low value below the maximum [Figs.\,\ref{figure4}\,(b) and (e)]. For all pressures studied, the temperature dependence observed under zfc/fh is perfectly consistent with the field-dependent data. 

The differences between the Hall signal recorded for fc/fh and zfc/fh are shown in Figs.\,\ref{figure4}\,(c) and (f). Panels on the left-hand side show data for a sample with a high RRR of 150, while the panels on the right-hand side show data for a low-quality sample with RRR of 45. For the low-quality sample, we observe much less variation for different pressures, while the size of the metastable Hall contribution in the high-quality sample varies a fair amount. This suggests that pinning at defects is needed to stabilize a metastable state of matter. Most remarkably, however, at the lowest temperatures the metastable signal contribution for all pressures and samples studied appears to limit around a similarly large value of $50\,{\rm n\Omega cm}$. In turn, the metastable Hall contribution offers the possibility to determine the size of the top-hat Hall signal without the effects of finite temperature (see Sec.\,\ref{discussion}).

\begin{figure*}
\includegraphics[width=0.95\textwidth]{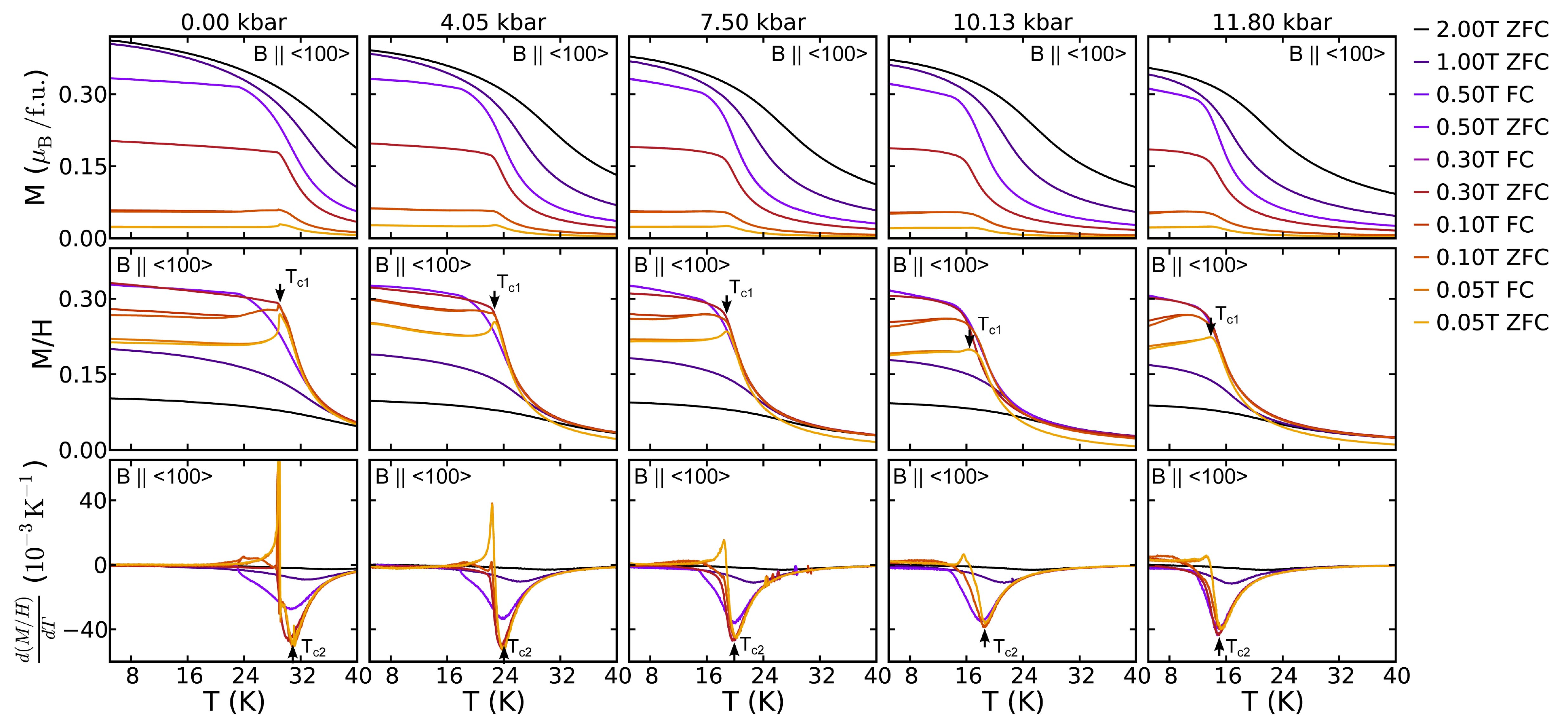}
\caption{(Color online) Magnetization as a function of temperature at various magnetic fields and pressures. Panels in the second row show the magnetization divided by the applied magnetic field to reveal better relative variations. Panels in the bottom row show the derivative $1/H {\rm d}M/{\rm d}T$. Data were recorded while heating in an applied field after field cooling or zero-field cooling, denoted as fc/fh and zfc/fh, respectively. Essentially, no difference is observed between fc/fh and zfc/fh.
}
\label{figure6}
\end{figure*}

To distinguish if the metastable signal contribution represents an anomalous or a topological Hall effect, we have measured the temperature dependence of the magnetization following the same field and temperature history. Typical data for various magnetic fields are shown in Fig.\,\ref{figure6}, where we find no difference under fc and zfc/fh. Panels in the first row of Fig.\,\ref{figure6} show the magnetization as a function of temperature as measured experimentally. In order to reveal better qualitative differences for the applied fields, we show in the second row of Fig.\,\ref{figure6} the ratio $M/H$. This highlights the absence of significant differences between zfc/fh and fc/fh magnetization data at all fields. The third row of Fig.\,\ref{figure6} displays, finally, the derivative of the magnetization with respect to the temperature. On the one hand, this permits to determine the transition temperature accurately. On the other hand, this corresponds to the magnetocaloric effect $dM/dT=dS/dB$. While the sharp spike near $T_c$ for $p=0$ vanishes with increasing pressure, we find essentially no changes of the qualitative behavior under applied fields. The reduction of the spike is thereby most likely the results of small pressure inhomogeneities across the sample volume. 

Taken together, the magnetization as a function of temperature clearly supports the interpretation that the metastable Hall signal represents a topological Hall signal. This  is also reflected by the temperature dependence of the Hall resistivity {\rxy} outside the field range of the top-hat contribution. Consistent with the anomalous Hall signal observed as a function of field, the Hall resistivity becomes very small due to the temperature dependence of {\rxx}, i.e., here the Hall signal corresponds to an intrinsic anomalous Hall conductivity that tracks the magnetization \cite{Lee:PRB07}. 


\subsection{Magnetic phase diagram}
\label{phasediagram}

Shown in Fig.\,\ref{figure7} are typical magnetic phase diagrams inferred from the magnetotransport data. The phase diagram at ambient pressure [Fig.\,\ref{figure7}\,(a)] is based  on susceptibility data as reported elsewhere, where we confirmed consistency with the features in our magnetotransport data (cf. Fig.\,\ref{figure13}; see also Ref.\cite{Neubauer:PRL2009}). With increasing pressure, the helimagnetic transition is suppressed, while the magnetic phase diagram does not change qualitatively. In particular, the lower critical field {\bco} appears to increase slightly under pressure. The small increase differs from small-angle neutron scattering (SANS) studies under pressure, where no pressure dependence was observed. However, we believe that the small increase of {\bco} under pressure inferred from the magnetoresistance reflects mostly changes of the form of  the magnetoresistance as discussed above. This may be compared with the upper critical field {\bct}, which does not change under pressure consistent with the SANS data. 

\begin{figure}
\includegraphics[width=0.3\textwidth]{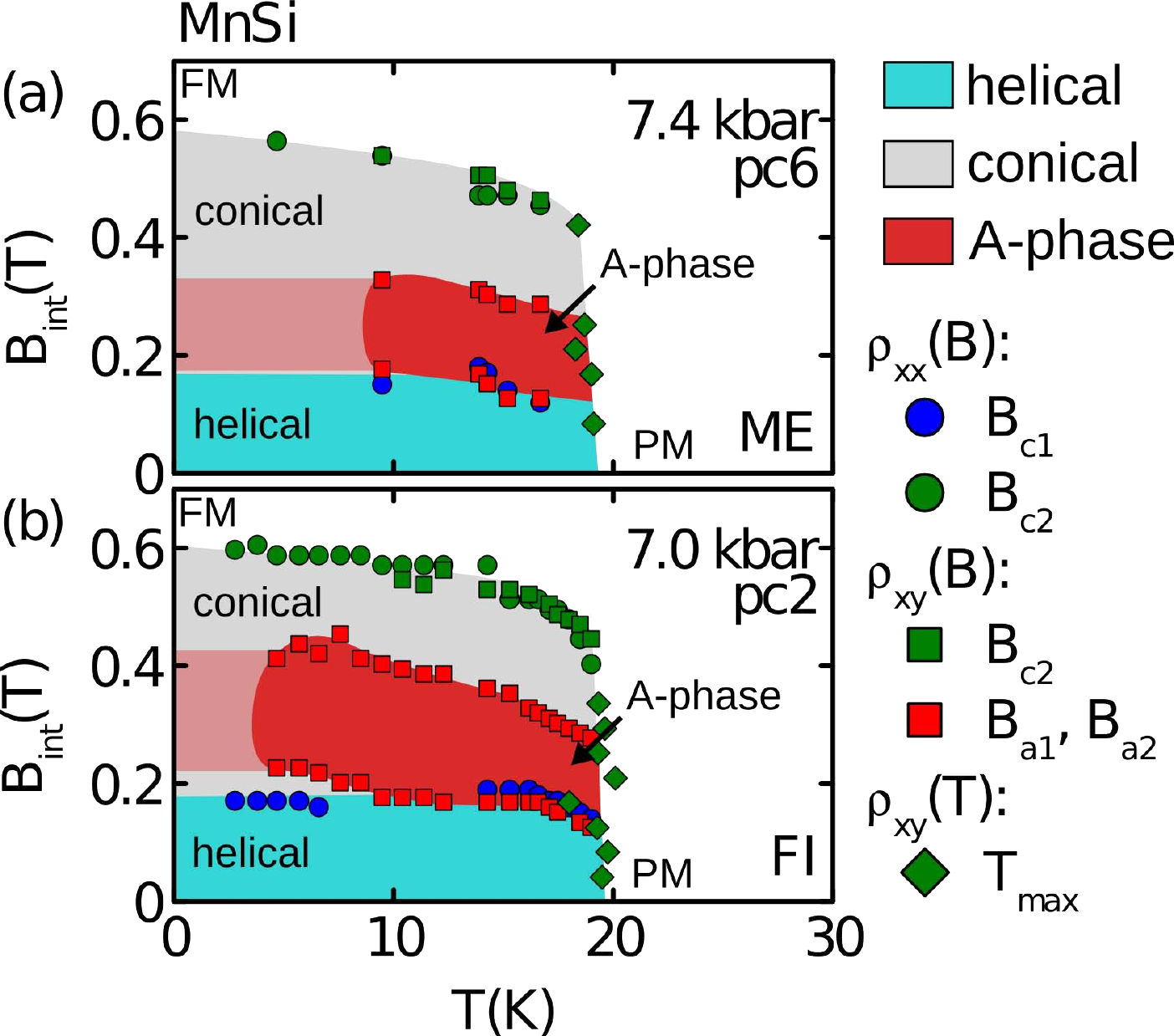}
\caption{(Color online) Magnetic phase diagram at $\sim7\,{\rm kbar}$ for different pressure transmitters. (a) Phase diagram inferred from data recorded with a methanol:ethanol (ME) mixture on a sample with RRR$=40$. (b) Phase diagram inferred from data recorded with a Fluorinert (FI) mixture on a sample with RRR$=$92. Fluorinert is known to be much less isotropic.
}
\label{figure8}
\end{figure}

The weak pressure dependence of {\bco} and {\bct} is contrasted by the regime in which a reversible top-hat (topological) Hall contribution (red shading) is observed. This comprises mostly data obtained in field sweeps. In comparison to ambient pressure, the field and temperature range increases under pressure. Yet, in contrast to the data reported in Ref.\,\cite{Lee:PRL09}, we still find a well-defined phase pocket that is strongly reminiscent of that seen at ambient pressure. This links the top-hat Hall contribution to the skyrmion lattice phase at $p=0$. As discussed below, our study even links the quantitative size of the top-hat signal to the skyrmion lattice phase at $p=0$. While the skyrmion lattice phase (denoted as A phase) increases in size, the only truly different property with respect to ambient pressure concerns the metastable behavior we observed under field cooling. This metastable behavior is indicated in the phase diagrams in terms of a light red shading extending down to the lowest temperatures studied.

As the metastable behavior emerges under pressure alongside the increase of the A phase, this raises the question to what extent it may be driven by pressure inhomogeneities and local strains. We have therefore also determined the magnetic phase diagram at a pressure around 7\,kbar for the methanol:ethanol and the Fluorinert mixture as pressure transmitter. As shown in Fig.\,\ref{figure8}, the extent of the A phase is considerably larger for the Fluorinert mixture. In combination with the sensitivity to sample purity, this appears consistent with the evidence reported elsewhere of less uniform pressure conditions for the Fluorinert mixture. However, typical anisotropies reported elsewhere of order $\lesssim 10^{-1}\,{\rm kbar}$ are tiny as compared with the overall pressure range of order $\sim 10\,{\rm kbar}$, i.e., a few \% (pressure inhomogeneities have been inferred, for instance, in Larmor diffraction \cite{Pfleiderer:Science07}).

\begin{figure}
\includegraphics[width=0.3\textwidth]{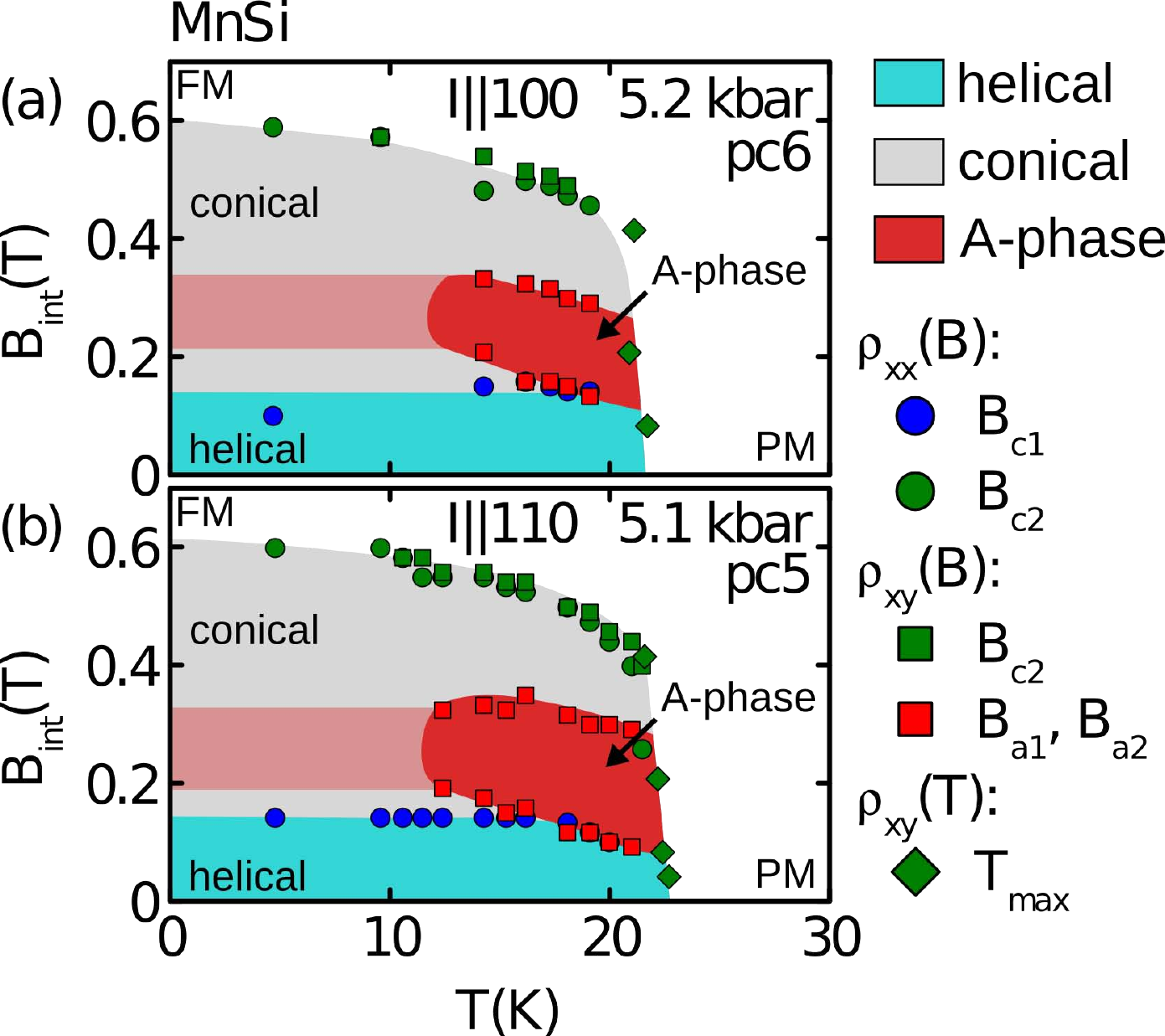}
\caption{(Color online) Magnetic phase diagram inferred from for the magnetotransport properties and two different current directions for $B\parallel${\ooz} at a pressure around $\sim 5\,{\rm kbar}$. (a) Phase diagram for current parallel {\ozz} on a sample with RRR$=$40. (b) Magnetic phase diagram for current parallel {\ooz} on a sample with RRR$=$46.
}
\label{figure9}
\end{figure}

A comparison of the magnetic phase diagrams for field parallel to {\ooz} and current along {\ozz} and {\ooz} is finally shown in Figs.\,\ref{figure9}\,(a) and \ref{figure9}\,(b), respectively. Within the accuracy of determining the phase boundaries, no differences are observed. This corresponds to the expected behavior, notably that the magnetic phases as inferred from the Hall signal are not sensitive to current direction.

\section{Discussion}

The discussion of our experimental results is organized in two parts. In the first part of this section we present theoretical aspects how the intrinsic anomalous Hall effect and the topological Hall effect as well as other factors determine the size of the Hall signal. This is followed by a discussion of the consistency of the experimental results with the theoretical description.

\subsection{Theory of the interplay of the topological and anomalous Hall effects}
\label{theory}

As emphasized in the introduction, the intrinsic anomalous Hall effect arises from Berry phases which an electron picks up when moving in momentum space, while the motion of the electron in the presence of a smooth magnetic texture is described by real-space Berry phases. In the following, we present a rather qualitative introduction of these effects. A full account of the complex interplay of various Berry phase terms in phase space will be developed in a future publication.

In a magnetic metal without inversion symmetry, the orientation of the spin of the electron is determined by two factors. First, the magnetism leads to an exchange splitting of the bands which can be described by a Zeeman field $\vec b^{\rm ex}$. By virtue of this field, the electron spins are aligned parallel or antiparallel to the magnetic field. Second, spin-orbit coupling in addition leads to a splitting of bands and the orientation of the spin becomes locked to its momentum. For weak spin-orbit coupling and smooth magnetic textures (as in MnSi), one can add up the two terms to obtain for a single band model the Hamiltonian
\begin{eqnarray}
H=\epsilon_{\vec p} \mathbbm{1}
+\vec g^{\rm SO}(\vec p) {\boldsymbol \sigma} + 
\vec b^{\rm ex}(\vec r)   {\boldsymbol \sigma},
\end{eqnarray}
where $\epsilon_{\vec p} \mathbbm{1}$ represents the band without the effects of exchange splitting and spin-orbit coupling. Further, we measure $\vec b^{\rm ex}$ in units of $| g \mu_B/2|=|\hbar g e/(4 m)|$ where $g$ is the $g$ factor, $e$ the electron charge, and $m$ the electron mass. The sign in the above equation takes into account that $e<0$. Therefore, the spin (magnetic moment) of an electron orients preferentially antiparallel (parallel) to the Zeeman field, respectively.

In the semiclassical limit, this allows to define the direction of the local magnetization $\hat n$, which is a function defined in the six-dimensional phase space comprising both position and momentum
\begin{eqnarray}
\hat{\vec n}(\vec x)=\frac{\vec g^{\rm SO}(\vec p)+\vec b^{\rm ex}(\vec r) }
{\left|\vec g^{\rm SO}(\vec p)+\vec b^{\rm ex}(\vec r)\right|}.
\end{eqnarray}
Here, we consider only situations, where as in MnSi the exchange fields vary on length scales much longer than the Fermi wave length. With $\downarrow$ we denote in the following a spin-orientation antiparallel to $\hat n$. As this is the spin orientation with the lower energy, a $\downarrow$ spin is carried by the majority electrons. Further, the coordinate in phase space is denoted as $\vec x = (\vec r,\vec p)$.

Berry phase effects induced by the change of the local wave function $|u(\vec x)\rangle$ of the majority spin $\downarrow$, defined by $[\hat{\vec n}(\vec x) {\boldsymbol \sigma}]\, |u(\vec x)\rangle=- |u(\vec x)\rangle$, are described by the six components of a Berry vector potential,
$q^e_\downarrow A_j(x)=i \hbar \langle u(\vec x)|\frac{\partial}{\partial x_j} |u(\vec x)\rangle$, $j=1,...,6$. Majority electrons, with a spin antiparallel to $\hat{\vec n}$ pick up the opposite Berry phase compared to minority electrons with parallel orientation. This is taken into account by attributing the charges $q^e_\downarrow=1/2$ and  $q^e_\uparrow=-1/2$ to the majority and minority electrons, respectively. Note that we use different sign conventions compared  to Ref.~\cite{Schulz:NaturePhysics2012}, correcting a typo in that paper.

The resulting effective magnetic fields are described by the antisymmetric $6 \times 6$ matrix \cite{Xiao:RMP10} 
\begin{eqnarray}
\Omega_{ij}=\frac{\partial A_j}{\partial x_i} -\frac{\partial A_i}{\partial x_j}= 
\hbar \hat{\vec n} \cdot \left(\frac{\partial}{\partial x_i} \hat{\vec n} \times \frac{\partial}{\partial x_j} \hat{\vec n}  \right).
\end{eqnarray}
The geometric interpretation of this term is that $\Omega_{ij}d x_i d x_j$ describes the Berry phase (times $\hbar$) picked up upon moving on an infinitesimal loop in the ($ij$) plane in phase space with area $dx_i dx_j$, which is given by the solid angle enclosed by the vectors $\hat{\vec n}(\vec x)$ in this loop. The first $3 \times 3$ components of the antisymmetric matrix $\boldsymbol \Omega$ are identified with the three components of the emergent magnetic field \cite{Schulz:NaturePhysics2012} arising from the real-space Berry phases only:
\begin{eqnarray}
B^{e}_i(\vec x)=\left| \frac{e}{ q^e_\sigma}  \right| B^{\rm eff}_i(\vec x)=\frac{1}{2} \sum_{j,k=1...3} \epsilon_{ijk} \Omega_{jk}
\end{eqnarray}
with $i \in \{1,2,3\}$. Note that the sum runs only over the real-space indices $1...3$. The emergent $B$ field $B^e$ has the units of $\hbar$ per area and is related
to $B^{\rm eff}$ (measured in tesla) used in Eq.~(\ref{beff}) by the factor $| e /q_\sigma^e |$.

Three other components ($i,j \in \{4,5,6\}$) of  $\Omega_{ij}$ describe the corresponding Berry phase fields in momentum space which are responsible for the intrinsic anomalous Hall effect \cite{Nagaosa:RMP10}. The remaining nine independent components of $\boldsymbol \Omega$ keep track of Berry phases picked up for loops in phase space involving both position and momentum directions which also contribute to the Hall effect.

The Berry fields $\boldsymbol \Omega$ determine the semiclassical equations of motion \cite{Xiao:RMP10}, 
\begin{eqnarray}
\partial_t x_i = J_{ij} (\frac{\partial \epsilon}{  \partial x_j}-q^e \Omega_{jk} \partial_t x_k)
\end{eqnarray}
or, equivalently, 
\begin{eqnarray}
(q^e_\sigma \vec \Omega-\vec J) \partial_t \vec x = \frac{\partial \epsilon}{  \partial \vec x},
\end{eqnarray}
where $\epsilon$ is the energy and 
\begin{eqnarray}
\vec J=\left( \begin{array}{cc} 0 &  \mathbbm{1} \\ - \mathbbm{1} & 0 \end{array}\right)
\end{eqnarray}
is the symplectic version of the identity ($\mathbbm{1}$ is the $3 \times 3$ identity matrix). Note that chiral metals with skyrmion lattices may be one of the first experimental systems where not only real-space and momentum-space Berry phases, but also mixed phase space Berry phases, described by the $6 \times 6$ matrix $\vec \Omega$, may become important. A full discussion of the corresponding contributions to the Hall effect is deferred to a future publication, while we focus in the following on the topological contribution.

When integrating  $\vec B^{e}$ over a magnetic unit cell of the skyrmion lattice (in real space), one obtains $\hbar$ times the total solid angle covered by $\hat{\vec n}$:
\begin{eqnarray}
\Phi^{e}(\vec p)&=&
\int_{\rm UC} \vec B^{e}(\vec x) \, \vec{d}^2 \vec{r}=\frac{2 \pi \hbar}{|q^e_\sigma|} \, n(\vec p) \\
 &=& \left\{ \begin{array}{cl}
0 \quad &  |\vec g^{\rm SO}(\vec{p})|>|\vec b^{\rm ex}(\vec r_0)|  \\
-\frac{2 \pi \hbar}{|q^e_\sigma|}  \quad & |\vec g^{\rm SO}(\vec{p})|<|\vec b^{\rm ex}(\vec r_0)| 
\end{array} \right. \label{flux}
\end{eqnarray}
Due to the periodic boundary condition, the total solid angle has to be a multiple of $4 \pi=2 \pi/|q^e_\sigma|$ which can be identified with a quantum of emergent flux. Therefore, $n(\vec p)$ is an integer. The topological winding number of the spin in real space determines directly the number of flux quanta per unit cell. As in the skyrmion lattice phase $\vec b^{\rm ex}(\vec r)$ winds once around the unit sphere with winding number $-1$, one obtains the flux $-\hbar 4 \pi$ when the exchange field $\vec B^{\rm ex}$ is larger than the spin splitting due to spin-orbit interactions. In the other limit, when $\vec g^{\rm SO}(\vec{p})$ is much larger than the the exchange field, the spin orientation within the unit cell only wiggles around its dominant direction  $\vec g^{\rm SO}(\vec{p})$ and the winding number vanishes. For fixed momentum $\vec p$, the transition from winding number $-1$ to $0$ occurs when at the point $\vec r_0$, where $\vec B^{\rm ex}(\vec r_0)$ is antiparallel to $\vec g^{\rm SO}(\vec{p})$, the two vectors compensate each other exactly, $\vec B^{\rm ex}(\vec r_0)+\vec g^{\rm SO}(\vec{p})=0$, such that locally the two bands cross.

In the limit $|\vec g^{\rm SO}| \ll |\vec b^{\rm ex}|$, one can ignore the spin-orbit coupling effects in the band structure. In this limit, the contribution to the Hall effect can be estimated from the Boltzmann equation using the relaxation-time approximation with spin-dependent relaxation time $\tau_{\downarrow}$ and $\tau_{\uparrow}$ for majority and minority spins, respectively. For $\vec B^e$ in the $z$-direction and $\vec k$-independent scattering rates, for example, the relaxation-time approximation predicts the following topological contribution: 
\begin{eqnarray}\label{sigmaxy}
\sigma^{\rm top}_{xy}&\approx & B^e    
\sum_{\sigma n}
 \int e^2 q^e_\sigma  \tau_{\sigma n}^2 
 \left(\frac{ (v_{\vec k n }^y)^2  }{m^{xx}_{\vec k n}} - \frac{ v_{\vec k n }^x v_{\vec k n }^y}{m^{xy}_{\vec k n}}\right) \notag\\
 &&\qquad\qquad\qquad \times\frac{\partial f_0(\epsilon_{\vec k \sigma n})}{\partial \epsilon} \frac{d^3 k}{(2 \pi)^3},
\end{eqnarray}
where $\vec v_{\vec k n}$ is the velocity in band $n$ and $m^{ij}_{\vec k n}=(\partial^2 \epsilon_{\vec k \sigma n}/\hbar^2\partial k_i \partial k_j)^{-1}$ are the elements of the effective mass tensor. One obtains exactly the same formulas for the Hall conductivity due to orbital magnetic fields, if one replaces the emergent charge $q^e_\sigma$ by the electron charge $e<0$. Therefore, it is convenient to express the topological Hall resistivity by the normal Hall coefficient $R_0$:
\begin{eqnarray}\label{rhoxy}
\rho^{\rm top}_{yx}&\approx &  R_0 B^e  \left\langle \frac{q^e_\sigma}{e} \right\rangle_{\rm FS} =   R_0 B^{\rm eff} P,
\end{eqnarray}
where
\begin{eqnarray}\label{P}
P&=&\left|\frac{e}{q^e_\sigma}\right|  \left\langle \frac{q^e_\sigma}{e} \right\rangle_{\rm FS}
\end{eqnarray}
is an effective polarization and
\begin{eqnarray}
\langle ... \rangle_{\rm FS}=\frac{\sum_{n\sigma} \int ... w_{\vec k \sigma n}}{\sum_{n\sigma} \int  w_{\vec k \sigma n}}
\end{eqnarray}
a certain average over all Fermi surfaces weighted by the square of the spin-dependent scattering rates, 
\begin{eqnarray}
w_{\vec k \sigma n}=\tau_{\sigma n}^2  \left(\frac{ (v_{\vec k n }^y)^2  }{m^{xx}_{\vec k n}} - \frac{ v_{\vec k n }^x v_{\vec k n }^y}{m^{xy}_{\vec k n}}\right) \frac{\partial f_0(\epsilon_{\vec k \sigma n})}{\partial \epsilon}.
\end{eqnarray}
In MnSi, the size of $B^{\rm eff}$ is given by Eq.~(\ref{beff}).

It is useful to discuss the sign of  $\left\langle q^e_\sigma/e \right\rangle_{\rm FS}$ and therefore of the effective polarization $P$. If the Fermi surface is electron like, the average is dominated by the majority spin with $q^e_\downarrow=1/2$ and the ratio  $\left\langle q^e_\sigma/e \right\rangle_{\rm FS}$ and therefore also $P$ is negative as the electron charge is negative, $e<0$. In contrast, for a hole like Fermi surface, we expect a higher density of states for minority spins and therefore $\left\langle q^e_\sigma/e \right\rangle_{\rm FS}>0$ and $P>0$. Because $B^e$ is antiparallel to the applied magnetic field in MnSi and the sign of the normal Hall effect suggests dominant holelike Fermi surfaces, one expects that the topological and normal contribution to the Hall effect have opposite sign consistent with experiment~\cite{Neubauer:PRL2009}. Note, however, that these simple rules can be violated in multiband systems due to the complicated Fermi-surface average.

As a final remark in this section, we note that one can repeat the same considerations also in the language of holes. Under a particle-hole transformation ($c^\dagger_\sigma \to \sigma c^\dagger_{-\sigma}$), the charge $e$ and the mass $m$ change sign, but the spin operator is not affected. Since a missing spin-up electron is a spin-down hole, an up-spin electron Fermi surface with emergent charge $q^e_\uparrow=-1/2$ maps to a spin-down hole Fermi surface with opposite emergent charge $q^e_\downarrow=1/2$. As above, we obtain that for a holelike Fermi surface $\left\langle q^e_\sigma/e \right\rangle_{\rm FS}$ is positive.

\subsection{Comparison with experiment}
\label{discussion}

In the light of the theoretical aspects presented above, we address now the following questions: Which factors determine  the size of the topological contribution to the Hall effect in MnSi? Can the giant contributions at finite pressures observed in MnSi be explained by the topologically quantized emergent field of the skyrmion lattice at ambient pressure? What is the generic size of the topological Hall contribution at low temperatures? Why is the topological signal at ambient pressure so much smaller?

In Fig.\,\ref{figure10}(a), we summarize two key features of the pressure dependence of the topological Hall resistivity we observe in our experiments. Shown by full symbols is the estimated maximum size of the top-hat-shaped topological Hall signal just below $T_c$, denoted {\Drxyt}, as measured under reversible conditions. This corresponds to data obtained in field sweeps (cf. Fig.\,\ref{figure3}), where a rough estimate of the anomalous Hall contribution is subtracted as illustrated in Fig.\,\ref{figure12}. In principle, samples with high and low RRRs, which display this maximum signal contribution at slightly different reduced temperatures as described above, show the same trends. With increasing pressure, the size of the revisable maximum signal increases, where the curve provides an estimated upper boundary as a guide to the eye. It is important to note that increasing pressure corresponds to decreasing {\tc} and thus decreasing absolute temperatures at which this signal is determined. The increase of {\Drxyt} with pressure hence corresponds also to an increase with decreasing temperature. 

To obtain an estimate of the generic value of the topological Hall signal in the low-temperature limit, we consider the metastable topological Hall contribution observed under field cooling. The open symbols in Fig.~\ref{figure10}\,(a) show the difference between the Hall signal observed under field cooling and zero-field cooling for a temperature of 2\,K. In this plot, for clarity only data of low RRR samples are shown since the metastable behavior is more pronounced for them. Yet, regardless of the RRR, the estimated zero-temperature contribution limits for a given pressure to the same low-temperature value [cf. Fig.\ref{figure4}\,(f)]. With increasing pressure, the extrapolated value of {\Drxyt} at 2\,K decreases. For pressures exceeding $p^*\sim12\,{\rm kbar}$, the pressure dependence of the open and filled symbols appears to merge (as emphasised above data for $p>p^*$ will be presented elsewhere \cite{Ritz:Nature2013} since the phase diagram displays further complexities above $p^*$ beyond the scope of the work presented here). We attribute the merging of the two pressure dependencies to the reduction of {\tc}, causing that both signals overlap significantly.  

It is now instructive to address the pressure and temperature dependencies of all factors entering equation (\ref{rhoxy}), namely, the size of the emergent field $B^e$, the normal Hall constant $R_0$, and the Fermi surface average $\langle q^e_\sigma/e \rangle_{\rm FS}$ of the emergent charge. 

We begin with the emergent field $B^e$. For low temperatures and deep in the ordered phase, we may assume that the size of the exchange splitting $\vec b^{\rm ex}$ is much larger than the spin-orbit splitting of the bands. Therefore, if the skyrmion lattice is unchanged, the average value of the emergent magnetic field $B^e$ is also unchanged and given by one flux quantum per magnetic unit cell. Unfortunately, a direct measure of the size of the magnetic unit cell of the skyrmion lattice phase as a function of pressure is, to our knowledge, presently not available. However, detailed neutron scattering studies of the helimagnetic order in MnSi have long established that the periodicity of the helix is essentially unchanged under pressure \cite{Pfleiderer:Nature2004,Pfleiderer:PRL07,Fak:JPCM2005}. This is also reflected by the lack of pressure dependence of {\bct}. Moreover, preliminary uniaxial pressure studies clearly show that the periodicity of the magnetic modulation remains unchanged while the alignment and orientation of the magnetic modulation responds sensitively to uniaxial stress \cite{Chacon:2012}. Finally, all B20 compounds studied to date which order helimagnetically display also a skyrmion lattice phase, where the magnetic periodicities in the skyrmion lattice phase and the helimagnetic state are consistent with each other and theory \cite{Muehlbauer:Science2009}. We therefore conclude that the magnetic unit cell must be essentially unchanged in size. Unless the topological winding number changes, which appears extremely unlikely, the strength of $B^e$ is almost pressure independent. The value of the emergent magnetic field is therefore essentially unchanged, $B^{\rm eff}\sim$-13.15\,T [cf. Eq.~(\ref{beff})] \cite{pressure}.


We address next the pressure dependence of the normal Hall constant $R_0$. A value of $R_0$ at low temperatures and high magnetic fields inferred from the experimental data is shown in Fig.~\ref{figure10}\,(b). $R_0$ is thereby approximately independent of pressure at high fields. Similar values for $R_0$ have been estimated from high-field data close to $T_c$ at ambient pressure \cite{Lee:PRB07,Neubauer:Diploma}.  A much more difficult question concerns variations of the normal Hall constant under changes of the size of the local magnetization at small fields and/or temperatures in the vicinity of $T_c$. Due to the large size of the anomalous Hall contribution, a reliable experimental determination of the normal Hall constant proves to be very difficult in this regime. More importantly, perhaps, the rather smooth field dependence observed experimentally does not indicate any particular complexities. 

In contrast, nonrelativistic band-structure calculations \cite{Jeong:PRB2004} suggest that the local magnetization may be accounted for by a rigid shift of minority versus majority bands as the local magnetization increases. Moreover, the calculated band structure suggests that the topology, shape, size, and sign of the effective masses of the minority and majority Fermi surfaces in MnSi change strongly when the magnetization increases from $0$ to $0.4\,\mu_B$, the size of the ordered moment at low $T$. Extensive de Haas-van Alphen measurements by Brown \cite{PhD-Brown} support this view as they reproduce in remarkable detail the experimentally observed Fermi surfaces at large magnetic fields. We therefore conclude that on the one hand, $R_0$ at fixed magnetization depends weakly on pressure. However, we cannot rule out definitively sizeable changes of $R_0$ for small values of the magnetization close to the transition temperature.

The remaining factor in Eq.~(\ref{rhoxy}) to be addressed is the Fermi-surface average of the emergent charge $\langle q^e_\sigma/e \rangle_{\rm FS}$. As $q^e_\sigma$ is of opposite sign for majority and minority spins, it may be expected to scale linearly with the strength of the local spin polarization in the absence of strong changes in the band structure, discussed in the following.  Shown by the straight line in Fig.\,\ref{figure10}\,(a) is a linear regression of the open symbols. The slope of this line corresponds to the relative pressure dependence of the magnetization ${\rm d}\,m_{\rm red}/{\rm d}p\approx-0.02\,{\rm kbar^{-1}}$ with $m_{\rm red}=m(p)/m(p=0)$, as extrapolated for zero field from fields above {\bct} \cite{Koyama:PRB00}. This rate of decrease is also consistent with the decrease of the ordered magnetic moment inferred from SANS \cite{Pfleiderer:Nature2004,Pfleiderer:PRL07}. The rate of decrease of {\Drxyt} with increasing pressure hence follows \textit{quantitatively} the pressure dependence of the spin polarization as expected from Eq.~(\ref{rhoxy}) and the weak pressure dependence of $B^e$ and $R_0$ discussed above. This provides further support of our interpretation of the topological Hall signal and of Eq.~(\ref{rhoxy}). 

Using Eq.~(\ref{rhoxy}), we obtain for the effective polarization
\begin{eqnarray}
P=\left| \frac{e}{q^e_\sigma}\right| \left\langle \frac{q^e_\sigma}{e} 
 \right\rangle_{\rm FS}  \approx \frac{\rho_{\rm yx}^{\rm top}}{R_0 B^{\rm eff}} \approx 0.22
\end{eqnarray}
at low temperatures and pressures where we used $\rho_{\rm yx}^{\rm top}\approx -50\,{\rm n\Omega\,cm}$, $B^{\rm eff}\approx -13.15\,{\rm T}$, and $R_0=1.7\cdot10^{-10}\,{\rm \Omega m T^{-1}}$ [for a value of $\lambda_{\rm S}\approx 180\,{\rm \AA}$ the polarisation is $P\approx0.27$ (Refs.~\cite{wavelength} and \cite{pressure})]. This is a reasonable value which is of the same order of magnitude as  the ratio $P_0=\mu_{\rm spo}/\mu_{\rm sat}\approx 0.18$ of the local magnetization (about $\mu_{\rm spo} \approx 0.4\,\mu_B$) and the nominal saturation moment of the Mn ions ($\mu_{\rm sat}\approx 2.2 \pm 0.2\,\mu_B$). Taking into account the complicated Fermi-surface averages determining $P$ and the complex band structure of MnSi \cite{Jeong:PRB2004,PhD-Brown}, this shows that the large signal at low temperatures and finite pressure can naturally be explained by the topological Hall effect arising from the skyrmion lattice.

The remaining question concerns the reduction of the topological signal at higher temperatures and, most importantly, the nature of the reduction of the signal size by about a factor $10$ at ambient pressure close to $T_c$. Here, one important factor is the reduction of the ordered moment (and therefore of $P$) when the temperature increases. In Ref.~\cite{Neubauer:PRL2009}, it was originally estimated that at ambient pressure the local polarization $P_0\approx 0.1$ is only a factor $2$ smaller than at $T=0$. While quantitative measurements of the size of the ordered moment in the skyrmion phase at ambient and high  pressure are presently not available due to subtle extinction effects, it is plausible that the reduction of the magnetization under pressure explains the decrease of the topological Hall signal with increasing pressure. 

Yet, even though the reduction of $P_0$ close to $T_c$ at ambient pressure was probably underestimated in Ref.~\cite{Neubauer:PRL2009}, we do not think that the linear dependence of  $\langle q^e_\sigma /e \rangle_{\rm FS}$ on $P_0$ fully explains the order-of-magnitude reduction of the topological Hall signal. 
Taking into account that an abundance of particle-hole excitations characteristic of the itinerant-electron magnetism of MnSi has been inferred from polarized neutron scattering \cite{Semadeni:PhysicaB99} as well as the temperature dependence of the damping of magnetic resonance data \cite{Schwarze:2012}, five mechanisms and combinations thereof may be at the heart of this reduction. 

First, as explained above, without the effects of scattering, one expects from band-structure calculations strong changes of the Fermi surface for variations at small values of the local magnetizations. This would modify $R_0$ and also the Fermi-surface averages $\langle q^e_\sigma /e \rangle_{\rm FS}$. Second, close to $T_c$, the relevant scattering processes may be completely different compared to the low-temperature situation. This can also strongly affect the Fermi-surface average  $\langle q^e_\sigma /e \rangle_{\rm FS}$ as different Fermi surfaces are weighted by the square of the scattering time. 

Third, as the exchange splitting close to $T_c$ is weak, it is possible that the spin-orbit splitting of the bands prohibits that the electron spin follows fully the magnetic texture. As shown in Eq.~(\ref{flux}), a possible consequence is that for some of the bands the topological contribution is completely switched off! It is also possible that only a part of a given Fermi surface is affected by the real-space Berry phase, while for other parts the emergent magnetic flux vanishes. Fourth, it is possible that close to $T_c$ the adiabatic approximation breaks down. Especially, if the spin-flip scattering length describing the scattering from minority to majority electrons (and vice versa) is smaller than the distance of the skyrmions, the topological Hall signal will become strongly suppressed. It is possible that the third and forth mechanisms are related. Namely, when spin-orbit splitting and exchange splitting are of similar magnitude, the splitting of majority and minority bands vanishes on three-dimensional planes in the six-dimensional phase space which may give rise to enhanced spin-flip scattering. 

Fifth and final, when spin-orbit and exchange splitting are of similar magnitude, one can not neglect the fact that aside from the real-space Berry phases, also the intrinsic anomalous Hall effect, caused by momentum-space Berry phases, is affected by the presence of the skyrmions and even new phase-space Berry phases emerge. These effects will be studied in the future. At present, neither the size nor the sign of these extra contributions are known theoretically or experimentally.

Most likely a combination of several of the effects described above is responsible for the strong reduction of the topological Hall contribution at ambient pressure. However, the experiments under pressure and the theoretical analysis show unambiguously that the giant low-temperature value of the topological signal is robustly given by a value of the order of 50 n$\Omega$cm, which depends only weakly on sample quality and pressure.

\section{Conclusions}
\label{conclusions}

In conclusion, we reported comprehensive measurements of the Hall effect in MnSi at low temperatures and high pressures across the magnetic phase diagram that reveal a large generic topological Hall signal. Exploring carefully the importance of the field and temperature history for the topological Hall signal opens an unexpected route to determine its generic size. Notably, tracking the topological Hall signal under field cooling allows essentially to switch off the effects of finite temperatures. Exploring the importance of the sample purity and pressure transmitter allows us to attribute the wide field range of the large topological Hall signal reported by Lee \textit{et al.} \cite{Lee:PRL09} to defect-induced pinning and pressure inhomogeneities. The field dependence observed in our study under improved experimental conditions thereby unambiguously links the large topological Hall signal to the skyrmion lattice phase at ambient pressure. 

As the large topological Hall signal clearly evolves under pressure out of the skyrmion lattice phase at ambient pressure, we can directly link it to the topological Hall signal arising from the winding of magnetization characteristic for skyrmion textures. It increases by about a factor of $10$ from the small signal observed at ambient pressure close to $T_c$ whenever the skyrmion lattice phase is stabilized at lower temperatures. The reduction of the ambient pressure signal arises very likely from a combination of several factors where the substantial reduction of the local polarization close to $T_c$ and associated changes of the Fermi surfaces are probably the most important ones. The size of the topological Hall signal at low temperature is, as in the case of the normal Hall effect, determined by how the scattering rates average over the various Fermi surfaces in MnSi.  Taken together, the increase of the topological Hall resistivity with increasing pressure (and hence decreasing helimagnetic transition temperature) arises clearly from a rather unusual combination of mechanisms. It is, nevertheless, fully compatible with the present understanding of the spin order and electronic properties of MnSi.\\

\acknowledgments 
We wish to thank P. B\"oni, K. Everschor, F. Freimuth, M. Garst, Y. Mokrousov, A. Neubauer, P.\,G. Niklowitz, M.~Janoschek, W. M\"unzer, and B. Russ for support and stimulating discussions. R.R., M.H., C.F., A.B., and M.W. acknowledge support through the TUM Graduate School. Financial support through DFG TRR80, SFB608, and FOR960 as well as ERC-AdG (291079 TOPFIT) are gratefully acknowledged. 


\begin{thebibliography}{66}
\expandafter\ifx\csname natexlab\endcsname\relax\def\natexlab#1{#1}\fi
\expandafter\ifx\csname bibnamefont\endcsname\relax
  \def\bibnamefont#1{#1}\fi
\expandafter\ifx\csname bibfnamefont\endcsname\relax
  \def\bibfnamefont#1{#1}\fi
\expandafter\ifx\csname citenamefont\endcsname\relax
  \def\citenamefont#1{#1}\fi
\expandafter\ifx\csname url\endcsname\relax
  \def\url#1{\texttt{#1}}\fi
\expandafter\ifx\csname urlprefix\endcsname\relax\def\urlprefix{URL }\fi
\providecommand{\bibinfo}[2]{#2}
\providecommand{\eprint}[2][]{\url{#2}}

\bibitem[{\citenamefont{Nagaosa et~al.}(2010)\citenamefont{Nagaosa, Sinova,
  Onoda, MacDonald, and Ong}}]{Nagaosa:RMP10}
\bibinfo{author}{\bibfnamefont{N.}~\bibnamefont{Nagaosa}},
  \bibinfo{author}{\bibfnamefont{J.}~\bibnamefont{Sinova}},
  \bibinfo{author}{\bibfnamefont{S.}~\bibnamefont{Onoda}},
  \bibinfo{author}{\bibfnamefont{A.~H.} \bibnamefont{MacDonald}},
  \bibnamefont{and} \bibinfo{author}{\bibfnamefont{N.}~\bibnamefont{Ong}},
  \bibinfo{journal}{Rev. Mod. Phys.} \textbf{\bibinfo{volume}{82}},
  \bibinfo{pages}{1539} (\bibinfo{year}{2010}).

\bibitem[{\citenamefont{Xiao et~al.}(2010)\citenamefont{Xiao, Chang, and
  Niu}}]{Xiao:RMP10}
\bibinfo{author}{\bibfnamefont{D.}~\bibnamefont{Xiao}},
  \bibinfo{author}{\bibfnamefont{M.-C.} \bibnamefont{Chang}}, \bibnamefont{and}
  \bibinfo{author}{\bibfnamefont{Q.}~\bibnamefont{Niu}}, \bibinfo{journal}{Rev.
  Mod. Phys.} \textbf{\bibinfo{volume}{82}}, \bibinfo{pages}{1959}
  (\bibinfo{year}{2010}).

\bibitem[{\citenamefont{Lee et~al.}(2004)\citenamefont{Lee, Watauchi, Miller,
  Cava, and Ong}}]{Lee:Science2004}
\bibinfo{author}{\bibfnamefont{W.-L.} \bibnamefont{Lee}},
  \bibinfo{author}{\bibfnamefont{S.}~\bibnamefont{Watauchi}},
  \bibinfo{author}{\bibfnamefont{V.~L.} \bibnamefont{Miller}},
  \bibinfo{author}{\bibfnamefont{R.~J.} \bibnamefont{Cava}}, \bibnamefont{and}
  \bibinfo{author}{\bibfnamefont{N.~P.} \bibnamefont{Ong}},
  \bibinfo{journal}{Science} \textbf{\bibinfo{volume}{303}},
  \bibinfo{pages}{1647} (\bibinfo{year}{2004}).

\bibitem[{\citenamefont{Taguchi et~al.}(2001)\citenamefont{Taguchi, Oohara,
  Yoshizawa, Nagaosa, and Tokura}}]{Taguchi:Science01}
\bibinfo{author}{\bibfnamefont{Y.}~\bibnamefont{Taguchi}},
  \bibinfo{author}{\bibfnamefont{Y.}~\bibnamefont{Oohara}},
  \bibinfo{author}{\bibfnamefont{H.}~\bibnamefont{Yoshizawa}},
  \bibinfo{author}{\bibfnamefont{N.}~\bibnamefont{Nagaosa}}, \bibnamefont{and}
  \bibinfo{author}{\bibfnamefont{Y.}~\bibnamefont{Tokura}},
  \bibinfo{journal}{Science} \textbf{\bibinfo{volume}{291}},
  \bibinfo{pages}{2573} (\bibinfo{year}{2001}).

\bibitem[{\citenamefont{Machida et~al.}(2007)\citenamefont{Machida, Nakatsuji,
  Maeno, Tayama, Sakakibara, and Onoda}}]{Machida:PRL07}
\bibinfo{author}{\bibfnamefont{Y.}~\bibnamefont{Machida}},
  \bibinfo{author}{\bibfnamefont{S.}~\bibnamefont{Nakatsuji}},
  \bibinfo{author}{\bibfnamefont{Y.}~\bibnamefont{Maeno}},
  \bibinfo{author}{\bibfnamefont{T.}~\bibnamefont{Tayama}},
  \bibinfo{author}{\bibfnamefont{T.}~\bibnamefont{Sakakibara}},
  \bibnamefont{and} \bibinfo{author}{\bibfnamefont{S.}~\bibnamefont{Onoda}},
  \bibinfo{journal}{Phys.\ Rev.\ Lett.} \textbf{\bibinfo{volume}{98}},
  \bibinfo{pages}{057203} (\bibinfo{year}{2007}).

\bibitem[{\citenamefont{Ueland et~al.}({2012})\citenamefont{Ueland, Miclea,
  Kato, Ayala-Valenzuela, McDonald, Okazaki, Tobash, Torrez, Ronning,
  Movshovich et~al.}}]{Ueland:NC2012}
\bibinfo{author}{\bibfnamefont{B.~G.} \bibnamefont{Ueland}},
  \bibinfo{author}{\bibfnamefont{C.~F.} \bibnamefont{Miclea}},
  \bibinfo{author}{\bibfnamefont{Y.}~\bibnamefont{Kato}},
  \bibinfo{author}{\bibfnamefont{O.}~\bibnamefont{Ayala-Valenzuela}},
  \bibinfo{author}{\bibfnamefont{R.~D.} \bibnamefont{McDonald}},
  \bibinfo{author}{\bibfnamefont{R.}~\bibnamefont{Okazaki}},
  \bibinfo{author}{\bibfnamefont{P.~H.} \bibnamefont{Tobash}},
  \bibinfo{author}{\bibfnamefont{M.~A.} \bibnamefont{Torrez}},
  \bibinfo{author}{\bibfnamefont{F.}~\bibnamefont{Ronning}},
  \bibinfo{author}{\bibfnamefont{R.}~\bibnamefont{Movshovich}},
  \bibnamefont{et~al.}, \bibinfo{journal}{{Nature Communicationss}}
  \textbf{\bibinfo{volume}{{3}}} (\bibinfo{year}{{2012}}).

\bibitem[{\citenamefont{Ye et~al.}(1999)\citenamefont{Ye, Kim, Millis,
  Shraiman, Majumdar, and Tessanovic}}]{Ye:PRL99}
\bibinfo{author}{\bibfnamefont{J.}~\bibnamefont{Ye}},
  \bibinfo{author}{\bibfnamefont{Y.~B.} \bibnamefont{Kim}},
  \bibinfo{author}{\bibfnamefont{A.}~\bibnamefont{Millis}},
  \bibinfo{author}{\bibfnamefont{B.}~\bibnamefont{Shraiman}},
  \bibinfo{author}{\bibfnamefont{P.}~\bibnamefont{Majumdar}}, \bibnamefont{and}
  \bibinfo{author}{\bibfnamefont{Z.}~\bibnamefont{Tessanovic}},
  \bibinfo{journal}{Phys. Rev. Lett.} \textbf{\bibinfo{volume}{83}},
  \bibinfo{pages}{3737} (\bibinfo{year}{1999}).

\bibitem[{\citenamefont{Yanagihara and Salamon}(2002)}]{Yanagihara:PRL02}
\bibinfo{author}{\bibfnamefont{H.}~\bibnamefont{Yanagihara}} \bibnamefont{and}
  \bibinfo{author}{\bibfnamefont{M.~B.} \bibnamefont{Salamon}},
  \bibinfo{journal}{Phys. Rev. Lett.} \textbf{\bibinfo{volume}{89}},
  \bibinfo{pages}{187201} (\bibinfo{year}{2002}).

\bibitem[{\citenamefont{Baily and Salamon}(2005)}]{Baily:PRB05}
\bibinfo{author}{\bibfnamefont{S.~A.} \bibnamefont{Baily}} \bibnamefont{and}
  \bibinfo{author}{\bibfnamefont{M.~B.} \bibnamefont{Salamon}},
  \bibinfo{journal}{Phys. Rev. B} \textbf{\bibinfo{volume}{71}},
  \bibinfo{pages}{104407} (\bibinfo{year}{2005}).

\bibitem[{\citenamefont{Neubauer
  et~al.}(2009{\natexlab{a}})\citenamefont{Neubauer, Pfleiderer, Binz, Rosch,
  Ritz, Niklowitz, and B{\"o}ni}}]{Neubauer:PRL2009}
\bibinfo{author}{\bibfnamefont{A.}~\bibnamefont{Neubauer}},
  \bibinfo{author}{\bibfnamefont{C.}~\bibnamefont{Pfleiderer}},
  \bibinfo{author}{\bibfnamefont{B.}~\bibnamefont{Binz}},
  \bibinfo{author}{\bibfnamefont{A.}~\bibnamefont{Rosch}},
  \bibinfo{author}{\bibfnamefont{R.}~\bibnamefont{Ritz}},
  \bibinfo{author}{\bibfnamefont{P.~G.} \bibnamefont{Niklowitz}},
  \bibnamefont{and} \bibinfo{author}{\bibfnamefont{P.}~\bibnamefont{B{\"o}ni}},
  \bibinfo{journal}{Phys. Rev. Lett.} \textbf{\bibinfo{volume}{102}},
  \bibinfo{pages}{186602} (\bibinfo{year}{2009}{\natexlab{a}}).

\bibitem[{\citenamefont{Lee et~al.}(2009)\citenamefont{Lee, Kang, Onose,
  Tokura, and Ong}}]{Lee:PRL09}
\bibinfo{author}{\bibfnamefont{M.}~\bibnamefont{Lee}},
  \bibinfo{author}{\bibfnamefont{W.}~\bibnamefont{Kang}},
  \bibinfo{author}{\bibfnamefont{Y.}~\bibnamefont{Onose}},
  \bibinfo{author}{\bibfnamefont{Y.}~\bibnamefont{Tokura}}, \bibnamefont{and}
  \bibinfo{author}{\bibfnamefont{N.}~\bibnamefont{Ong}},
  \bibinfo{journal}{Phys. Rev. Lett.} \textbf{\bibinfo{volume}{102}},
  \bibinfo{pages}{186601} (\bibinfo{year}{2009}).

\bibitem[{\citenamefont{Jonietz et~al.}(2010)\citenamefont{Jonietz,
  M{\"u}hlbauer, Pfleiderer, Neubauer, M{\"u}nzer, Bauer, Adams, Georgii,
  B{\"o}ni, Duine et~al.}}]{Jonietz:Science2010}
\bibinfo{author}{\bibfnamefont{F.}~\bibnamefont{Jonietz}},
  \bibinfo{author}{\bibfnamefont{S.}~\bibnamefont{M{\"u}hlbauer}},
  \bibinfo{author}{\bibfnamefont{C.}~\bibnamefont{Pfleiderer}},
  \bibinfo{author}{\bibfnamefont{A.}~\bibnamefont{Neubauer}},
  \bibinfo{author}{\bibfnamefont{W.}~\bibnamefont{M{\"u}nzer}},
  \bibinfo{author}{\bibfnamefont{A.}~\bibnamefont{Bauer}},
  \bibinfo{author}{\bibfnamefont{T.}~\bibnamefont{Adams}},
  \bibinfo{author}{\bibfnamefont{R.}~\bibnamefont{Georgii}},
  \bibinfo{author}{\bibfnamefont{P.}~\bibnamefont{B{\"o}ni}},
  \bibinfo{author}{\bibfnamefont{R.~A.} \bibnamefont{Duine}},
  \bibnamefont{et~al.}, \bibinfo{journal}{Science}
  \textbf{\bibinfo{volume}{330}}, \bibinfo{pages}{1648} (\bibinfo{year}{2010}).

\bibitem[{\citenamefont{Schulz et~al.}(2012)\citenamefont{Schulz, Ritz, Bauer,
  Halder, Wagner, Franz, Pfleiderer, Everschor, Garst, and
  Rosch}}]{Schulz:NaturePhysics2012}
\bibinfo{author}{\bibfnamefont{T.}~\bibnamefont{Schulz}},
  \bibinfo{author}{\bibfnamefont{R.}~\bibnamefont{Ritz}},
  \bibinfo{author}{\bibfnamefont{A.}~\bibnamefont{Bauer}},
  \bibinfo{author}{\bibfnamefont{M.}~\bibnamefont{Halder}},
  \bibinfo{author}{\bibfnamefont{M.}~\bibnamefont{Wagner}},
  \bibinfo{author}{\bibfnamefont{C.}~\bibnamefont{Franz}},
  \bibinfo{author}{\bibfnamefont{C.}~\bibnamefont{Pfleiderer}},
  \bibinfo{author}{\bibfnamefont{K.}~\bibnamefont{Everschor}},
  \bibinfo{author}{\bibfnamefont{M.}~\bibnamefont{Garst}}, \bibnamefont{and}
  \bibinfo{author}{\bibfnamefont{A.}~\bibnamefont{Rosch}},
  \bibinfo{journal}{Nature Physics} \textbf{\bibinfo{volume}{8}},
  \bibinfo{pages}{301} (\bibinfo{year}{2012}).

\bibitem[{\citenamefont{Kanazawa et~al.}(2011)\citenamefont{Kanazawa, Onose,
  Arima, Okuyama, Ohoyama, Wakimoto, Kakurai, Ishiwata, and
  Tokura}}]{Kanzawa:PRL2011}
\bibinfo{author}{\bibfnamefont{N.}~\bibnamefont{Kanazawa}},
  \bibinfo{author}{\bibfnamefont{Y.}~\bibnamefont{Onose}},
  \bibinfo{author}{\bibfnamefont{T.}~\bibnamefont{Arima}},
  \bibinfo{author}{\bibfnamefont{D.}~\bibnamefont{Okuyama}},
  \bibinfo{author}{\bibfnamefont{K.}~\bibnamefont{Ohoyama}},
  \bibinfo{author}{\bibfnamefont{S.}~\bibnamefont{Wakimoto}},
  \bibinfo{author}{\bibfnamefont{K.}~\bibnamefont{Kakurai}},
  \bibinfo{author}{\bibfnamefont{S.}~\bibnamefont{Ishiwata}}, \bibnamefont{and}
  \bibinfo{author}{\bibfnamefont{Y.}~\bibnamefont{Tokura}},
  \bibinfo{journal}{Phys. Rev. Lett.} \textbf{\bibinfo{volume}{106}},
  \bibinfo{pages}{156603} (\bibinfo{year}{2011}).

\bibitem[{\citenamefont{Ishiwata et~al.}(2011)\citenamefont{Ishiwata, Tokunaga,
  Kaneko, Okuyama, Tokunaga, Wakimoto, Kakurai, Arima, Taguchi, and
  Tokura}}]{Ishiwata:PRB2011}
\bibinfo{author}{\bibfnamefont{S.}~\bibnamefont{Ishiwata}},
  \bibinfo{author}{\bibfnamefont{M.}~\bibnamefont{Tokunaga}},
  \bibinfo{author}{\bibfnamefont{Y.}~\bibnamefont{Kaneko}},
  \bibinfo{author}{\bibfnamefont{D.}~\bibnamefont{Okuyama}},
  \bibinfo{author}{\bibfnamefont{Y.}~\bibnamefont{Tokunaga}},
  \bibinfo{author}{\bibfnamefont{S.}~\bibnamefont{Wakimoto}},
  \bibinfo{author}{\bibfnamefont{K.}~\bibnamefont{Kakurai}},
  \bibinfo{author}{\bibfnamefont{T.}~\bibnamefont{Arima}},
  \bibinfo{author}{\bibfnamefont{Y.}~\bibnamefont{Taguchi}}, \bibnamefont{and}
  \bibinfo{author}{\bibfnamefont{Y.}~\bibnamefont{Tokura}},
  \bibinfo{journal}{Phys. Rev. B} \textbf{\bibinfo{volume}{84}},
  \bibinfo{pages}{054427} (\bibinfo{year}{2011}).

\bibitem[{\citenamefont{Landau and Lifshitz}(1980)}]{Landau}
\bibinfo{author}{\bibfnamefont{L.~D.} \bibnamefont{Landau}} \bibnamefont{and}
  \bibinfo{author}{\bibfnamefont{E.~M.} \bibnamefont{Lifshitz}},
  \emph{\bibinfo{title}{Course of theoretical physics, vol. 8}}
  (\bibinfo{publisher}{Pergamon Press}, \bibinfo{year}{1980}).

\bibitem[{wav()}]{wavelength}
\bibinfo{note}{The helical modulation increases from $\lambda_{\rm h}
  \sim\,180{\rm \AA}$ for $(T\to0)$ to $\lambda_{\rm h} \sim\,165{\rm \AA}$ for
  $T\to T_c$ (see, e.g., Refs.\,
  \cite{Ishikawa:SSC1976,Grigoriev:PRB2006,Janoschek:PRB2013}).}

\bibitem[{\citenamefont{Kusaka et~al.}(1976)\citenamefont{Kusaka, Yamamoto,
  Komatsubara, and Ishikawa}}]{Kusaka:SSC1976}
\bibinfo{author}{\bibfnamefont{S.}~\bibnamefont{Kusaka}},
  \bibinfo{author}{\bibfnamefont{K.}~\bibnamefont{Yamamoto}},
  \bibinfo{author}{\bibfnamefont{T.}~\bibnamefont{Komatsubara}},
  \bibnamefont{and} \bibinfo{author}{\bibfnamefont{Y.}~\bibnamefont{Ishikawa}},
  \bibinfo{journal}{Solid State Communications} \textbf{\bibinfo{volume}{20}},
  \bibinfo{pages}{925} (\bibinfo{year}{1976}).

\bibitem[{\citenamefont{Kadowaki et~al.}(1982)\citenamefont{Kadowaki, Okuda,
  and Date}}]{Kadowaki:JPSJ1982}
\bibinfo{author}{\bibfnamefont{K.}~\bibnamefont{Kadowaki}},
  \bibinfo{author}{\bibfnamefont{K.}~\bibnamefont{Okuda}}, \bibnamefont{and}
  \bibinfo{author}{\bibfnamefont{M.}~\bibnamefont{Date}}, \bibinfo{journal}{J.
  Phys. Soc. Jpn.} \textbf{\bibinfo{volume}{51}}, \bibinfo{pages}{2433}
  (\bibinfo{year}{1982}).

\bibitem[{\citenamefont{Bauer and Pfleiderer}(2012)}]{Bauer:PRB12}
\bibinfo{author}{\bibfnamefont{A.}~\bibnamefont{Bauer}} \bibnamefont{and}
  \bibinfo{author}{\bibfnamefont{C.}~\bibnamefont{Pfleiderer}},
  \bibinfo{journal}{Phys. Rev. B} \textbf{\bibinfo{volume}{85}},
  \bibinfo{pages}{214418} (\bibinfo{year}{2012}).

\bibitem[{\citenamefont{M\"uhlbauer et~al.}(2009)\citenamefont{M\"uhlbauer,
  Binz, Jonietz, Pfleiderer, Rosch, Neubauer, Georgii, and
  B\"oni}}]{Muehlbauer:Science2009}
\bibinfo{author}{\bibfnamefont{S.}~\bibnamefont{M\"uhlbauer}},
  \bibinfo{author}{\bibfnamefont{B.}~\bibnamefont{Binz}},
  \bibinfo{author}{\bibfnamefont{F.}~\bibnamefont{Jonietz}},
  \bibinfo{author}{\bibfnamefont{C.}~\bibnamefont{Pfleiderer}},
  \bibinfo{author}{\bibfnamefont{A.}~\bibnamefont{Rosch}},
  \bibinfo{author}{\bibfnamefont{A.}~\bibnamefont{Neubauer}},
  \bibinfo{author}{\bibfnamefont{R.}~\bibnamefont{Georgii}}, \bibnamefont{and}
  \bibinfo{author}{\bibfnamefont{P.}~\bibnamefont{B\"oni}},
  \bibinfo{journal}{Science} \textbf{\bibinfo{volume}{323}},
  \bibinfo{pages}{915} (\bibinfo{year}{2009}).

\bibitem[{\citenamefont{Adams et~al.}(2011)\citenamefont{Adams, M{\"u}hlbauer,
  Pfleiderer, Jonietz, Bauer, Neubauer, Georgii, B{\"o}ni, Keiderling,
  Everschor et~al.}}]{Adams:PRL2011}
\bibinfo{author}{\bibfnamefont{T.}~\bibnamefont{Adams}},
  \bibinfo{author}{\bibfnamefont{S.}~\bibnamefont{M{\"u}hlbauer}},
  \bibinfo{author}{\bibfnamefont{C.}~\bibnamefont{Pfleiderer}},
  \bibinfo{author}{\bibfnamefont{F.}~\bibnamefont{Jonietz}},
  \bibinfo{author}{\bibfnamefont{A.}~\bibnamefont{Bauer}},
  \bibinfo{author}{\bibfnamefont{A.}~\bibnamefont{Neubauer}},
  \bibinfo{author}{\bibfnamefont{R.}~\bibnamefont{Georgii}},
  \bibinfo{author}{\bibfnamefont{P.}~\bibnamefont{B{\"o}ni}},
  \bibinfo{author}{\bibfnamefont{U.}~\bibnamefont{Keiderling}},
  \bibinfo{author}{\bibfnamefont{K.}~\bibnamefont{Everschor}},
  \bibnamefont{et~al.}, \bibinfo{journal}{Phys. Rev. Lett.}
  \textbf{\bibinfo{volume}{107}}, \bibinfo{pages}{217206}
  (\bibinfo{year}{2011}).

\bibitem[{\citenamefont{Tonumura et~al.}(2012)\citenamefont{Tonumura, Yu,
  Yanagisawa, Matsuda, Onose, Kanazawa, Park, and
  Tokura}}]{Tonumura:NanoLetters2012}
\bibinfo{author}{\bibfnamefont{A.}~\bibnamefont{Tonumura}},
  \bibinfo{author}{\bibfnamefont{X.}~\bibnamefont{Yu}},
  \bibinfo{author}{\bibfnamefont{K.}~\bibnamefont{Yanagisawa}},
  \bibinfo{author}{\bibfnamefont{T.}~\bibnamefont{Matsuda}},
  \bibinfo{author}{\bibfnamefont{Y.}~\bibnamefont{Onose}},
  \bibinfo{author}{\bibfnamefont{N.}~\bibnamefont{Kanazawa}},
  \bibinfo{author}{\bibfnamefont{H.~S.} \bibnamefont{Park}}, \bibnamefont{and}
  \bibinfo{author}{\bibfnamefont{Y.}~\bibnamefont{Tokura}},
  \bibinfo{journal}{Nano Letters} \textbf{\bibinfo{volume}{12}},
  \bibinfo{pages}{1673} (\bibinfo{year}{2012}).

\bibitem[{com({\natexlab{a}})}]{comment-machida07}
\bibinfo{note}{The effective field defined in Ref.\,\cite{Machida:PRL07} for a
  discrete lattice of spins limits to the definition of the topological field
  for a continuous magnetization used here.}

\bibitem[{\citenamefont{Lee et~al.}(2007)\citenamefont{Lee, Onose, Tokura, and
  Ong}}]{Lee:PRB07}
\bibinfo{author}{\bibfnamefont{M.}~\bibnamefont{Lee}},
  \bibinfo{author}{\bibfnamefont{Y.}~\bibnamefont{Onose}},
  \bibinfo{author}{\bibfnamefont{Y.}~\bibnamefont{Tokura}}, \bibnamefont{and}
  \bibinfo{author}{\bibfnamefont{N.~P.} \bibnamefont{Ong}},
  \bibinfo{journal}{Phys. Rev. B} \textbf{\bibinfo{volume}{75}},
  \bibinfo{pages}{172403} (\bibinfo{year}{2007}).

\bibitem[{\citenamefont{Neubauer
  et~al.}(2009{\natexlab{b}})\citenamefont{Neubauer, Pfleiderer, Ritz,
  Niklowitz, and B{\"o}ni}}]{Neubauer:PhysicaB09}
\bibinfo{author}{\bibfnamefont{A.}~\bibnamefont{Neubauer}},
  \bibinfo{author}{\bibfnamefont{C.}~\bibnamefont{Pfleiderer}},
  \bibinfo{author}{\bibfnamefont{R.}~\bibnamefont{Ritz}},
  \bibinfo{author}{\bibfnamefont{P.~G.} \bibnamefont{Niklowitz}},
  \bibnamefont{and} \bibinfo{author}{\bibfnamefont{P.}~\bibnamefont{B{\"o}ni}},
  \bibinfo{journal}{Physica B} \textbf{\bibinfo{volume}{404}},
  \bibinfo{pages}{3163} (\bibinfo{year}{2009}{\natexlab{b}}).

\bibitem[{cla()}]{clarification}
\bibinfo{note}{The sign of the Hall signal reported by Lee \textit{et al.} in
  Refs.\,\cite{Lee:PRL09} and \cite{Lee:PRB07} as well as that given in
  Ref.\,\cite{Neubauer:PRL2009} is correct. The sign of the Hall signal
  reported in Ref.\,\cite{Neubauer:PhysicaB09} is not correct, contrary to what
  is stated in this paper.}

\bibitem[{\citenamefont{Bruno et~al.}(2004)\citenamefont{Bruno, Dugaev, and
  Taillefumier}}]{Bruno:PRL04}
\bibinfo{author}{\bibfnamefont{P.}~\bibnamefont{Bruno}},
  \bibinfo{author}{\bibfnamefont{V.~K.} \bibnamefont{Dugaev}},
  \bibnamefont{and}
  \bibinfo{author}{\bibfnamefont{M.}~\bibnamefont{Taillefumier}},
  \bibinfo{journal}{Phys. Rev. Lett.} \textbf{\bibinfo{volume}{93}},
  \bibinfo{pages}{096806} (\bibinfo{year}{2004}).

\bibitem[{\citenamefont{Tatara et~al.}(2007)\citenamefont{Tatara, Kohno,
  Shibata, Lemaho, and Lee}}]{Tatara:JPSJ07}
\bibinfo{author}{\bibfnamefont{G.}~\bibnamefont{Tatara}},
  \bibinfo{author}{\bibfnamefont{H.}~\bibnamefont{Kohno}},
  \bibinfo{author}{\bibfnamefont{J.}~\bibnamefont{Shibata}},
  \bibinfo{author}{\bibfnamefont{Y.}~\bibnamefont{Lemaho}}, \bibnamefont{and}
  \bibinfo{author}{\bibfnamefont{K.-J.} \bibnamefont{Lee}},
  \bibinfo{journal}{J. Phys. Soc. Jpn.} \textbf{\bibinfo{volume}{76}},
  \bibinfo{pages}{054707} (\bibinfo{year}{2007}).

\bibitem[{\citenamefont{Binz and Vishwanath}(2008)}]{Binz:PhysicaB08}
\bibinfo{author}{\bibfnamefont{B.}~\bibnamefont{Binz}} \bibnamefont{and}
  \bibinfo{author}{\bibfnamefont{A.}~\bibnamefont{Vishwanath}},
  \bibinfo{journal}{Physica B} \textbf{\bibinfo{volume}{403}},
  \bibinfo{pages}{1336} (\bibinfo{year}{2008}).

\bibitem[{err()}]{erratum}
\bibinfo{note}{An erratum to Ref.\,\cite{Neubauer:PRL2009} has been submitted.}

\bibitem[{\citenamefont{Pfleiderer et~al.}(1997)\citenamefont{Pfleiderer,
  McMullan, Julian, and Lonzarich}}]{Pfleiderer:PRB97}
\bibinfo{author}{\bibfnamefont{C.}~\bibnamefont{Pfleiderer}},
  \bibinfo{author}{\bibfnamefont{G.~J.} \bibnamefont{McMullan}},
  \bibinfo{author}{\bibfnamefont{S.~R.} \bibnamefont{Julian}},
  \bibnamefont{and} \bibinfo{author}{\bibfnamefont{G.~G.}
  \bibnamefont{Lonzarich}}, \bibinfo{journal}{Phys. Rev. B}
  \textbf{\bibinfo{volume}{55}}, \bibinfo{pages}{8330} (\bibinfo{year}{1997}).

\bibitem[{\citenamefont{Thessieu et~al.}(1995)\citenamefont{Thessieu, Flouquet,
  Lapertot, Stepanov, and Jaccard}}]{Thessieu:SSC95}
\bibinfo{author}{\bibfnamefont{C.}~\bibnamefont{Thessieu}},
  \bibinfo{author}{\bibfnamefont{J.}~\bibnamefont{Flouquet}},
  \bibinfo{author}{\bibfnamefont{G.}~\bibnamefont{Lapertot}},
  \bibinfo{author}{\bibfnamefont{A.~N.} \bibnamefont{Stepanov}},
  \bibnamefont{and} \bibinfo{author}{\bibfnamefont{D.}~\bibnamefont{Jaccard}},
  \bibinfo{journal}{Solid State Communications} \textbf{\bibinfo{volume}{95}},
  \bibinfo{pages}{707} (\bibinfo{year}{1995}).

\bibitem[{\citenamefont{Doiron-Leyraud
  et~al.}(2003)\citenamefont{Doiron-Leyraud, Walker, Taillefer, Steiner,
  Julian, and Lonzarich}}]{Doiron:Nature03}
\bibinfo{author}{\bibfnamefont{N.}~\bibnamefont{Doiron-Leyraud}},
  \bibinfo{author}{\bibfnamefont{I.~R.} \bibnamefont{Walker}},
  \bibinfo{author}{\bibfnamefont{L.}~\bibnamefont{Taillefer}},
  \bibinfo{author}{\bibfnamefont{M.~J.} \bibnamefont{Steiner}},
  \bibinfo{author}{\bibfnamefont{S.~R.} \bibnamefont{Julian}},
  \bibnamefont{and} \bibinfo{author}{\bibfnamefont{G.~G.}
  \bibnamefont{Lonzarich}}, \bibinfo{journal}{Nature}
  \textbf{\bibinfo{volume}{425}}, \bibinfo{pages}{595} (\bibinfo{year}{2003}).


\bibitem[{\citenamefont{Petrova
  et~al.}(2012)\citenamefont{Petrova, Stishov}}]{Petrova:PRB2012}
\bibinfo{author}{\bibfnamefont{A.~E.}~\bibnamefont{Petrova}},
  \bibnamefont{and} \bibinfo{author}{\bibfnamefont{S.~M.} \bibnamefont{Stishov}},
  \bibinfo{journal}{Phys. Rev. B}
  \textbf{\bibinfo{volume}{86}}, \bibinfo{pages}{174407} (\bibinfo{year}{2012}).


\bibitem[{\citenamefont{Thessieu et~al.}(1997)\citenamefont{Thessieu,
  Pfleiderer, Stepanov, and Flouquet}}]{Thessieu:JPCM97}
\bibinfo{author}{\bibfnamefont{C.}~\bibnamefont{Thessieu}},
  \bibinfo{author}{\bibfnamefont{C.}~\bibnamefont{Pfleiderer}},
  \bibinfo{author}{\bibfnamefont{A.~N.} \bibnamefont{Stepanov}},
  \bibnamefont{and} \bibinfo{author}{\bibfnamefont{J.}~\bibnamefont{Flouquet}},
  \bibinfo{journal}{J. Phys.: Cond. Matter} \textbf{\bibinfo{volume}{9}},
  \bibinfo{pages}{6677} (\bibinfo{year}{1997}).

\bibitem[{\citenamefont{Bloch et~al.}(1975)\citenamefont{Bloch, Voiron,
  Jaccarino, and Wernick}}]{Bloch:PLA74}
\bibinfo{author}{\bibfnamefont{D.}~\bibnamefont{Bloch}},
  \bibinfo{author}{\bibfnamefont{J.}~\bibnamefont{Voiron}},
  \bibinfo{author}{\bibfnamefont{V.}~\bibnamefont{Jaccarino}},
  \bibnamefont{and} \bibinfo{author}{\bibfnamefont{J.~H.}
  \bibnamefont{Wernick}}, \bibinfo{journal}{Phys. Lett. A}
  \textbf{\bibinfo{volume}{51}}, \bibinfo{pages}{259} (\bibinfo{year}{1975}).

\bibitem[{\citenamefont{Thessieu et~al.}(1998)\citenamefont{Thessieu,
  Kamishima, Goto, and Lapertot}}]{Thessieu:JPSJ98}
\bibinfo{author}{\bibfnamefont{C.}~\bibnamefont{Thessieu}},
  \bibinfo{author}{\bibfnamefont{K.}~\bibnamefont{Kamishima}},
  \bibinfo{author}{\bibfnamefont{T.}~\bibnamefont{Goto}}, \bibnamefont{and}
  \bibinfo{author}{\bibfnamefont{G.}~\bibnamefont{Lapertot}},
  \bibinfo{journal}{J. Phys. Soc. Jpn.} \textbf{\bibinfo{volume}{67}},
  \bibinfo{pages}{3605} (\bibinfo{year}{1998}).

\bibitem[{\citenamefont{Koyama et~al.}(2000)\citenamefont{Koyama, Goto,
  Kanomata, and Note}}]{Koyama:PRB00}
\bibinfo{author}{\bibfnamefont{K.}~\bibnamefont{Koyama}},
  \bibinfo{author}{\bibfnamefont{T.}~\bibnamefont{Goto}},
  \bibinfo{author}{\bibfnamefont{T.}~\bibnamefont{Kanomata}}, \bibnamefont{and}
  \bibinfo{author}{\bibfnamefont{R.}~\bibnamefont{Note}},
  \bibinfo{journal}{Phys. Rev. B} \textbf{\bibinfo{volume}{62}},
  \bibinfo{pages}{986} (\bibinfo{year}{2000}).

\bibitem[{\citenamefont{Pfleiderer
  et~al.}(2007{\natexlab{a}})\citenamefont{Pfleiderer, B\"oni, Keller,
  R\"o{\ss}ler, and Rosch}}]{Pfleiderer:Science07}
\bibinfo{author}{\bibfnamefont{C.}~\bibnamefont{Pfleiderer}},
  \bibinfo{author}{\bibfnamefont{P.}~\bibnamefont{B\"oni}},
  \bibinfo{author}{\bibfnamefont{T.}~\bibnamefont{Keller}},
  \bibinfo{author}{\bibfnamefont{U.~K.} \bibnamefont{R\"o{\ss}ler}},
  \bibnamefont{and} \bibinfo{author}{\bibfnamefont{A.}~\bibnamefont{Rosch}},
  \bibinfo{journal}{Science} \textbf{\bibinfo{volume}{316}},
  \bibinfo{pages}{1871} (\bibinfo{year}{2007}{\natexlab{a}}).

\bibitem[{\citenamefont{Cheng et~al.}(2010)\citenamefont{Cheng, Zhou, Zhou,
  Goodenough, and Sui}}]{Cheng:PRB10}
\bibinfo{author}{\bibfnamefont{J.-G.} \bibnamefont{Cheng}},
  \bibinfo{author}{\bibfnamefont{F.}~\bibnamefont{Zhou}},
  \bibinfo{author}{\bibfnamefont{J.-S.} \bibnamefont{Zhou}},
  \bibinfo{author}{\bibfnamefont{J.~B.} \bibnamefont{Goodenough}},
  \bibnamefont{and} \bibinfo{author}{\bibfnamefont{Y.}~\bibnamefont{Sui}},
  \bibinfo{journal}{Phys. Rev. B} \textbf{\bibinfo{volume}{82}},
  \bibinfo{pages}{214402} (\bibinfo{year}{2010}).

\bibitem[{\citenamefont{Pfleiderer
  et~al.}(2004{\natexlab{a}})\citenamefont{Pfleiderer, Reznik, Pintschovius,
  v.~L\"ohn\-eysen, Garst, and Rosch}}]{Pfleiderer:Nature2004}
\bibinfo{author}{\bibfnamefont{C.}~\bibnamefont{Pfleiderer}},
  \bibinfo{author}{\bibfnamefont{D.}~\bibnamefont{Reznik}},
  \bibinfo{author}{\bibfnamefont{L.}~\bibnamefont{Pintschovius}},
  \bibinfo{author}{\bibfnamefont{H.}~\bibnamefont{v.~L\"ohn\-eysen}},
  \bibinfo{author}{\bibfnamefont{M.}~\bibnamefont{Garst}}, \bibnamefont{and}
  \bibinfo{author}{\bibfnamefont{A.}~\bibnamefont{Rosch}},
  \bibinfo{journal}{Nature} \textbf{\bibinfo{volume}{427}},
  \bibinfo{pages}{227} (\bibinfo{year}{2004}{\natexlab{a}}).

\bibitem[{\citenamefont{Pfleiderer
  et~al.}(2007{\natexlab{b}})\citenamefont{Pfleiderer, Reznik, Pintschovius,
  and Haug}}]{Pfleiderer:PRL07}
\bibinfo{author}{\bibfnamefont{C.}~\bibnamefont{Pfleiderer}},
  \bibinfo{author}{\bibfnamefont{D.}~\bibnamefont{Reznik}},
  \bibinfo{author}{\bibfnamefont{L.}~\bibnamefont{Pintschovius}},
  \bibnamefont{and} \bibinfo{author}{\bibfnamefont{J.}~\bibnamefont{Haug}},
  \bibinfo{journal}{Physical Review Letters} \textbf{\bibinfo{volume}{99}},
  \bibinfo{eid}{156406} (\bibinfo{year}{2007}{\natexlab{b}}).

\bibitem[{\citenamefont{Yu et~al.}(2004)\citenamefont{Yu, Zamborszky, Thompson,
  Sarrao, Torelli, Fisk, and Brown}}]{Yu:PRL04}
\bibinfo{author}{\bibfnamefont{W.}~\bibnamefont{Yu}},
  \bibinfo{author}{\bibfnamefont{F.}~\bibnamefont{Zamborszky}},
  \bibinfo{author}{\bibfnamefont{J.~D.} \bibnamefont{Thompson}},
  \bibinfo{author}{\bibfnamefont{J.~L.} \bibnamefont{Sarrao}},
  \bibinfo{author}{\bibfnamefont{M.~E.} \bibnamefont{Torelli}},
  \bibinfo{author}{\bibfnamefont{Z.}~\bibnamefont{Fisk}}, \bibnamefont{and}
  \bibinfo{author}{\bibfnamefont{S.~E.} \bibnamefont{Brown}},
  \bibinfo{journal}{Phys. Rev. Lett.} \textbf{\bibinfo{volume}{92}},
  \bibinfo{pages}{086403} (\bibinfo{year}{2004}).

\bibitem[{\citenamefont{Uemura et~al.}(2007)\citenamefont{Uemura, Goko,
  Gat-Malureanu, Carlo, Russo, Savici, Aczel, MacDougall, Rodoriguez, Luke
  et~al.}}]{Uemura:NaturePhysics07}
\bibinfo{author}{\bibfnamefont{Y.~J.} \bibnamefont{Uemura}},
  \bibinfo{author}{\bibfnamefont{T.}~\bibnamefont{Goko}},
  \bibinfo{author}{\bibfnamefont{I.~M.} \bibnamefont{Gat-Malureanu}},
  \bibinfo{author}{\bibfnamefont{J.~P.} \bibnamefont{Carlo}},
  \bibinfo{author}{\bibfnamefont{P.~L.} \bibnamefont{Russo}},
  \bibinfo{author}{\bibfnamefont{A.~T.} \bibnamefont{Savici}},
  \bibinfo{author}{\bibfnamefont{A.}~\bibnamefont{Aczel}},
  \bibinfo{author}{\bibfnamefont{G.~J.} \bibnamefont{MacDougall}},
  \bibinfo{author}{\bibfnamefont{J.}~\bibnamefont{Rodoriguez}},
  \bibinfo{author}{\bibfnamefont{G.~M.} \bibnamefont{Luke}},
  \bibnamefont{et~al.}, \bibinfo{journal}{Nature Physics}
  \textbf{\bibinfo{volume}{3}}, \bibinfo{pages}{34} (\bibinfo{year}{2007}).

\bibitem[{com({\natexlab{b}})}]{comment-stishov}
\bibinfo{note}{We note that the experimental results reported in
  Ref.\,\cite{Petrova:PRB2012} and references therein are consistent with all other
  studies published, being characteristic of a smearing out of the first-order
  behavior by inhomogeneities under pressure for the sample quality studied.
  The interpretation offered in this paper completely ignores the appearance of
  itinerant metamagnetism for $p>p^*$
  (Refs.\,\cite{Thessieu:JPCM97}, \cite{Koyama:PRB00}, and \cite{Pfleiderer:PRL07}) as the main argument in support of a first-order quantum phase transition.}


\bibitem[{\citenamefont{Ritz et~al.}(2013)\citenamefont{Ritz, Halder, Wagner, Franz, Bauer, Pfleiderer}}]{Ritz:Nature2013}
\bibinfo{author}{\bibfnamefont{R.}~\bibnamefont{Ritz}},
  \bibinfo{author}{\bibfnamefont{M.}~\bibnamefont{Halder}},
  \bibinfo{author}{\bibfnamefont{M.}~\bibnamefont{Wagner}},
  \bibinfo{author}{\bibfnamefont{C.}~\bibnamefont{Franz}},
  \bibinfo{author}{\bibfnamefont{A.}~\bibnamefont{Bauer}},
  \bibnamefont{and}
  \bibinfo{author}{\bibfnamefont{C.}~\bibnamefont{Pfleiderer}},
  \bibinfo{journal}{Nature} \textbf{\bibinfo{volume}{497}},
  \bibinfo{pages}{231--234} (\bibinfo{year}{2013}).


\bibitem[{\citenamefont{Neubauer et~al.}(2011)\citenamefont{Neubauer, B{\oe}uf,
  Bauer, Russ, v.~L\"ohneyen, and Pfleiderer}}]{Neubauer:RSI2011}
\bibinfo{author}{\bibfnamefont{A.}~\bibnamefont{Neubauer}},
  \bibinfo{author}{\bibfnamefont{J.}~\bibnamefont{B{\oe}uf}},
  \bibinfo{author}{\bibfnamefont{A.}~\bibnamefont{Bauer}},
  \bibinfo{author}{\bibfnamefont{B.}~\bibnamefont{Russ}},
  \bibinfo{author}{\bibfnamefont{H.}~\bibnamefont{v.~L\"ohneysen}},
  \bibnamefont{and}
  \bibinfo{author}{\bibfnamefont{C.}~\bibnamefont{Pfleiderer}},
  \bibinfo{journal}{Rev. Sci. Instrum.} \textbf{\bibinfo{volume}{82}},
  \bibinfo{pages}{013902} (\bibinfo{year}{2011}).

  
\bibitem[{\citenamefont{Bauer}(2014)}]{Bauer:PhD}
\bibinfo{author}{\bibfnamefont{A.}~\bibnamefont{Bauer}},
  \bibinfo{type}{forthcoming {PhD} thesis}, \bibinfo{school}{Technische Universit\"at
  M\"unchen} (\bibinfo{year}{2014}).

\bibitem[{\citenamefont{F{\aa}k et~al.}(2005)\citenamefont{F{\aa}k, Sadykov,
  Flouquet, and Lapertot}}]{Fak:JPCM2005}
\bibinfo{author}{\bibfnamefont{B.}~\bibnamefont{F{\aa}k}},
  \bibinfo{author}{\bibfnamefont{R.}~\bibnamefont{Sadykov}},
  \bibinfo{author}{\bibfnamefont{J.}~\bibnamefont{Flouquet}}, \bibnamefont{and}
  \bibinfo{author}{\bibfnamefont{G.}~\bibnamefont{Lapertot}},
  \bibinfo{journal}{J. Phys.: Condens. Matter} \textbf{\bibinfo{volume}{17}},
  \bibinfo{pages}{1635} (\bibinfo{year}{2005}).

\bibitem[{\citenamefont{Aharoni}(1998)}]{Aharoni:JAP98}
\bibinfo{author}{\bibfnamefont{A.}~\bibnamefont{Aharoni}}, \bibinfo{journal}{J.
  Appl. Phys.} \textbf{\bibinfo{volume}{83}}, \bibinfo{pages}{3432}
  (\bibinfo{year}{1998}).

\bibitem[{\citenamefont{Thomasson et~al.}(1990)\citenamefont{Thomasson, Ayache,
  Spain, and Villedieu}}]{Thomasson}
\bibinfo{author}{\bibfnamefont{J.}~\bibnamefont{Thomasson}},
  \bibinfo{author}{\bibfnamefont{C.}~\bibnamefont{Ayache}},
  \bibinfo{author}{\bibfnamefont{I.~L.} \bibnamefont{Spain}}, \bibnamefont{and}
  \bibinfo{author}{\bibfnamefont{M.}~\bibnamefont{Villedieu}},
  \bibinfo{journal}{J. Appl. Phys.} \textbf{\bibinfo{volume}{68}},
  \bibinfo{pages}{5933} (\bibinfo{year}{1990}).

\bibitem[{\citenamefont{Sidorov and Sadykov}(2005)}]{Sidorov:JPCM2005}
\bibinfo{author}{\bibfnamefont{V.}~\bibnamefont{Sidorov}} \bibnamefont{and}
  \bibinfo{author}{\bibfnamefont{R.}~\bibnamefont{Sadykov}},
  \bibinfo{journal}{J. Phys.: Condens. Matter} \textbf{\bibinfo{volume}{17}},
  \bibinfo{pages}{S3008} (\bibinfo{year}{2005}).

\bibitem[{\citenamefont{Pfleiderer
  et~al.}(2004{\natexlab{b}})\citenamefont{Pfleiderer, Huxley, and
  Hayden}}]{Pfleiderer:JPCM2004}
\bibinfo{author}{\bibfnamefont{C.}~\bibnamefont{Pfleiderer}},
  \bibinfo{author}{\bibfnamefont{A.}~\bibnamefont{Huxley}}, \bibnamefont{and}
  \bibinfo{author}{\bibfnamefont{S.}~\bibnamefont{Hayden}},
  \bibinfo{journal}{J. Phys.: Condens. Matter} \textbf{\bibinfo{volume}{17}},
  \bibinfo{pages}{S3111} (\bibinfo{year}{2004}{\natexlab{b}}),
  \bibinfo{note}{and references therein}.

\bibitem[{\citenamefont{M\"unzer et~al.}(2010)\citenamefont{M\"unzer, Neubauer,
  Adams, M\"uhlbauer, Franz, Jonietz, Georgii, B\"oni, Pedersen, Schmidt
  et~al.}}]{Muenzer:PRB2010}
\bibinfo{author}{\bibfnamefont{W.}~\bibnamefont{M\"unzer}},
  \bibinfo{author}{\bibfnamefont{A.}~\bibnamefont{Neubauer}},
  \bibinfo{author}{\bibfnamefont{T.}~\bibnamefont{Adams}},
  \bibinfo{author}{\bibfnamefont{S.}~\bibnamefont{M\"uhlbauer}},
  \bibinfo{author}{\bibfnamefont{C.}~\bibnamefont{Franz}},
  \bibinfo{author}{\bibfnamefont{F.}~\bibnamefont{Jonietz}},
  \bibinfo{author}{\bibfnamefont{R.}~\bibnamefont{Georgii}},
  \bibinfo{author}{\bibfnamefont{P.}~\bibnamefont{B\"oni}},
  \bibinfo{author}{\bibfnamefont{B.}~\bibnamefont{Pedersen}},
  \bibinfo{author}{\bibfnamefont{M.}~\bibnamefont{Schmidt}},
  \bibnamefont{et~al.}, \bibinfo{journal}{Phys. Rev. B (R)}
  \textbf{\bibinfo{volume}{81}}, \bibinfo{pages}{041203}
  (\bibinfo{year}{2010}).

\bibitem[{Cha()}]{Chacon:2012}
\bibinfo{note}{A. Chacon, T. Adams, G. Brandl, A. Bauer, R. Georgii, P.
  B{\"o}ni, C. Pfleiderer, Uniaxial stress studies of the skyrmion lattice in
  MnSi, unpublished (2012)}.

\bibitem[{pre()}]{pressure}
\bibinfo{note}{If the wavelength in the metastable regime follows that of the
  helical state, we expect an increase of $\lambda_{\rm S}$ by roughly 10\% for
  $T\to0$ that implies a reduction of the emergent field by roughy 20\,\% (Ref.\,\cite{wavelength}}).

\bibitem[{\citenamefont{Neubauer}(2006)}]{Neubauer:Diploma}
\bibinfo{author}{\bibfnamefont{A.}~\bibnamefont{Neubauer}},
  \bibinfo{type}{{Diploma} thesis}, \bibinfo{school}{Technische Universit\"at
  M\"unchen} (\bibinfo{year}{2006}).

\bibitem[{\citenamefont{Jeong and Pickett}(2004)}]{Jeong:PRB2004}
\bibinfo{author}{\bibfnamefont{T.}~\bibnamefont{Jeong}} \bibnamefont{and}
  \bibinfo{author}{\bibfnamefont{W.~E.} \bibnamefont{Pickett}},
  \bibinfo{journal}{Phys. Rev. B} \textbf{\bibinfo{volume}{70}},
  \bibinfo{pages}{075114} (\bibinfo{year}{2004}).

\bibitem[{\citenamefont{Brown}(1990)}]{PhD-Brown}
\bibinfo{author}{\bibfnamefont{S.~A.} \bibnamefont{Brown}},
  \bibinfo{type}{{PhD} thesis}, \bibinfo{school}{University of Cambridge}
  (\bibinfo{year}{1990}).


\bibitem[{\citenamefont{Semadeni et~al.}(1999)\citenamefont{Semadeni, B\"oni,
  Endoh, Roessli, and Shirane}}]{Semadeni:PhysicaB99}
\bibinfo{author}{\bibfnamefont{F.}~\bibnamefont{Semadeni}},
  \bibinfo{author}{\bibfnamefont{P.}~\bibnamefont{B\"oni}},
  \bibinfo{author}{\bibfnamefont{Y.}~\bibnamefont{Endoh}},
  \bibinfo{author}{\bibfnamefont{B.}~\bibnamefont{Roessli}}, \bibnamefont{and}
  \bibinfo{author}{\bibfnamefont{G.}~\bibnamefont{Shirane}},
  \bibinfo{journal}{Physica B} \textbf{\bibinfo{volume}{267-268}},
  \bibinfo{pages}{248} (\bibinfo{year}{1999}).

\bibitem[{Sch()}]{Schwarze:2012}
\bibinfo{note}{T. Schwarze, A. Bauer, H. Berger, J.Waizner, C. Pfleiderer, A.
  Rosch, D. Grundler, Ferromagnetic resonance measurements in helimagnetic B20
  compounds, unpublished}.

\bibitem[{\citenamefont{Ishikawa et~al.}(1976)\citenamefont{Ishikawa, Tajima,
  Bloch, and Roth}}]{Ishikawa:SSC1976}
\bibinfo{author}{\bibfnamefont{Y.}~\bibnamefont{Ishikawa}},
  \bibinfo{author}{\bibfnamefont{K.}~\bibnamefont{Tajima}},
  \bibinfo{author}{\bibfnamefont{D.}~\bibnamefont{Bloch}}, \bibnamefont{and}
  \bibinfo{author}{\bibfnamefont{M.}~\bibnamefont{Roth}},
  \bibinfo{journal}{Solid State Communications} \textbf{\bibinfo{volume}{19}},
  \bibinfo{pages}{525} (\bibinfo{year}{1976}).

\bibitem[{\citenamefont{Grigoriev et~al.}(2006)\citenamefont{Grigoriev,
  Maleyev, Okorokov, Chetverikov, B\"oni, Georgii, Lamago, Eckerlebe, and
  Pranzas}}]{Grigoriev:PRB2006}
\bibinfo{author}{\bibfnamefont{S.~V.} \bibnamefont{Grigoriev}},
  \bibinfo{author}{\bibfnamefont{S.~V.} \bibnamefont{Maleyev}},
  \bibinfo{author}{\bibfnamefont{A.~I.} \bibnamefont{Okorokov}},
  \bibinfo{author}{\bibfnamefont{Y.~O.} \bibnamefont{Chetverikov}},
  \bibinfo{author}{\bibfnamefont{P.}~\bibnamefont{B\"oni}},
  \bibinfo{author}{\bibfnamefont{R.}~\bibnamefont{Georgii}},
  \bibinfo{author}{\bibfnamefont{D.}~\bibnamefont{Lamago}},
  \bibinfo{author}{\bibfnamefont{H.}~\bibnamefont{Eckerlebe}},
  \bibnamefont{and} \bibinfo{author}{\bibfnamefont{K.}~\bibnamefont{Pranzas}},
  \bibinfo{journal}{Phys. Rev. B} \textbf{\bibinfo{volume}{74}},
  \bibinfo{pages}{214414} (\bibinfo{year}{2006}).

  
\bibitem[{\citenamefont{Janoschek et~al.}(2013)\citenamefont{Janoschek, Garst, Bauer, Krautscheid, Georgii, B\"oni, Pfleiderer}}]{Janoschek:PRB2013}
\bibinfo{author}{\bibfnamefont{M.} \bibnamefont{Janoschek}},
  \bibinfo{author}{\bibfnamefont{M.} \bibnamefont{Garst}},
  \bibinfo{author}{\bibfnamefont{A.} \bibnamefont{Bauer}},
  \bibinfo{author}{\bibfnamefont{P.} \bibnamefont{Krautscheid}},
  \bibinfo{author}{\bibfnamefont{R.} \bibnamefont{Georgii}},
  \bibinfo{author}{\bibfnamefont{P.} \bibnamefont{B\"oni}},
  \bibnamefont{and} \bibinfo{author}{\bibfnamefont{C.} \bibnamefont{Pfleiderer}},
  \bibinfo{journal}{Phys. Rev. B} \textbf{\bibinfo{volume}{87}},
  \bibinfo{pages}{134407} (\bibinfo{year}{2013}).


\end{thebibliography}


\end{document}